\documentclass[apj,twocolumn,appendixfloats,twocolappendix]{openjournal}

\usepackage[T1]{fontenc}
\usepackage[utf8]{inputenc}

\usepackage{amsmath} 
\usepackage{amsfonts} 
\usepackage{amsbsy} 
\usepackage{amssymb} 

\usepackage{graphicx} 

\usepackage{hyperref} 

\usepackage{acronym} 
\usepackage{soul} 
\usepackage{xcolor} 

\usepackage{orcidlink} 

\hypersetup{
    colorlinks=true,
    linkcolor=blue,
    filecolor=blue,
    urlcolor=blue,
    citecolor=blue,
}

\newcommand{\pushright}[1]{\tag*{$#1$}}

\newacro{PPD}{proto-planetary disc}

\newacroplural{PPD}{proto-planetary discs}

\newacro{BC}{boundary condition}
\newcommand{\BC}{\ac{BC}}
\newacroplural{BC}{boundary conditions}
\newcommand{\BCs}{\acp{BC}}

\newacro{PDE}{partial differential equation}

\newacro{ODE}{ordinary differential equation}
\newcommand{\ODE}{\ac{ODE}}

\newacro{RWI}{Rossby wave instability}
\newcommand{\RWI}{\ac{RWI}}

\newcommand{\bb}[1]{\mbox{\boldmath{$#1$}}}

\newcommand{\rd}{\mathrm{d}} 

\newcommand{\bcdot}{\, \mbox{\boldmath{$\cdot$}} \,}

\newcommand{\ex}{\ensuremath{\bb{e}_{x}}} 
\newcommand{\ey}{\ensuremath{\bb{e}_{y}}} 
\newcommand{\er}{\ensuremath{\bb{e}_{r}}} 

\newcommand{\br}{\mathbf{r}} 
\newcommand{\ba}{\mathbf{a}} 

\newcommand{\e}{\mathrm{e}} 

\begin{document}

\title{Reflex instabilities II: The reflex motion of the central object drives at least seven instabilities in accretion discs}
\shorttitle{Reflex instabilities II}

\author{\vspace{-40pt}Nathan Magnan \orcidlink{0000-0001-8602-9725}$^{1, 2}$}
\author{Michel Tagger$^{3}$}
\author{Clément Baruteau$^{4}$}
\author{Aurélien Crida$^{1}$}
\author{Jean-François Gonzalez \orcidlink{0000-0001-9423-6062}$^{5}$}
\author{Héloïse Méheut$^{1}$}
\affiliation{$^{1}$ Université Côte d'Azur, Observatoire de la Côte d'Azur, CNRS, Laboratoire Lagrange, France}
\affiliation{$^{2}$ Centre National d’\'Etudes Spatiales (CNES), Paris, France}
\affiliation{$^{3}$ LPC2E, OSUC, Observatoire de Paris, Université d'Orléans, Université Paris Sciences et Lettres, CNRS, CNES, Orléans, France}
\affiliation{$^{4}$ IRAP, Université de Toulouse, CNRS, CNES, Toulouse, France}
\affiliation{$^{5}$ Université Lyon 1, ENS de Lyon, CNRS, CRAL, UMR 5574, Saint-Genis-Laval, France
\vspace{+0.5 \baselineskip}
}
\email{Corresponding author: nathan.magnan@oca.eu}

\begin{abstract}
    This series of papers studies the stability of accretion discs to the reflex motion of the central object under the gravitational pull of the disc itself. In this second paper, we use linear analysis to reproduce or discover seven distinct instabilities. We outline their defining characteristics and propose a classification. Of the four instabilities that had already been discovered, three are artificial: one does not survive the transition to modern surface density profiles such as power-laws, one is due to the location of the domain's boundaries, and one is due to the use of damping layers near the domain's boundaries. Only the SLING instability of \cite{Adams+89} and the three new `cavity' instabilities seem capable of affecting accretion discs. All four can be described to a high degree of precision by WKB methods. They may drive eccentricity growth, affect planet migration, create spiral and crescent substructures in the absence of any planet, and boost accretion onto the central object. The issue is that the three artificial instabilities are extremely hard to avoid in simulations, and would bias the same observables. This suggests that our community may not have the tools to model the reflex motion of the central object yet.
\end{abstract}

\maketitle

\section{Introduction}
\label{sec:Introduction}

Many accretion discs show asymmetric features, including eccentricity \citep{Rees88, Fragile+26} and large-scale spirals and crescents \citep{Bae+23}. The dipolar (\textit{i.e.} \smash{${m = 1}$}) component of those asymmetries can gravitationally displace the central object, thereby modifying the gravitational field experienced by the disc. This creates a feedback loop whose stability has been debated since the late eighties \citep{Adams+89, Shu+90, Noh+92, Ostriker+92} and has recently attracted renewed attention \citep{Crida+25b, Sanchez+25}.

The confusing thing is that each of those research groups reports a different disc behaviour. For instance, the modes found by \cite{Adams+89} rely on an outer Lindblad resonance, those found by \cite{Heemskerk+92} are located between two Lindblad resonances, and those found by \cite{Crida+25b} are retrograde. This all seems contradictory, but the point of the present paper is to show that the system does indeed support several instabilities. The aforementionned teams found four different ones, and we report the discovery of three more.

Our paper also aims to explore and organise this diversity. It is structured as follows: we introduce the governing equations in \S\ref{sec:Governing_equations}, describe the equilibrium state of the system in \S\ref{sec:Equilibrium}, derive the linear perturbation equations in \S\ref{sec:Linear_perturbation_equations}, and explain our parameter space exploration strategy in \S\ref{sec:Exploration_strategy}. We then study the `bracket' modes in \S\ref{sec:Bracket_modes}, three groups of `cavity' modes in \S\ref{sec:Cavity_modes}, two groups of self-gravity-induced modes in \S\ref{sec:Gravitational_modes}, and `overdamped' modes in \S\ref{sec:Overdamped_modes}. The meanings of those names are explained in the corresponding sections. Finally, we discuss the consequences of our findings in~\S\ref{sec:Discussion_conclusion}.

An important piece of context to keep in mind is that \{central object + accretion disc\} systems are prone to artificial linear modes. For instance, \cite{Sanchez+25}'s modes are caused by the location of the edges of the simulation domain. Even worse, we showed in our first paper that their number and stability depend on the choice of reference frame (stellocentric or barycentric) and the choice of equations for the central object's motion (Newton's law or centre-of-mass conservation). Similar issues affect all the modes discussed here.

\section{Governing equations}
\label{sec:Governing_equations}

\newcommand{\Ms}{M_{\star}} 
\newcommand{\Md}{M_{d}} 

We consider a two-dimensional fluid disc of mass $\Md$ around a point-like central object of mass $\Ms$. We introduce the equations that govern the disc's evolution in \S\ref{sub:Governing_equations_disc}, those governing the central object's motion in \S\ref{sub:Governing_equations_central_object}, and those modelling self-gravity in~\S\ref{sub:Governing_equations_self_gravity}.\footnote{Note that in order to follow the \href{https://textrecycling.org/resources/text-recycling-policy/}{TRRP's guidelines}, \S\ref{sec:Governing_equations}, \S\ref{sec:Equilibrium} and~\S\ref{sec:Linear_perturbation_equations} are copies of the corresponding sections of paper 1.}

\subsection{Governing equations for the disc}
\label{sub:Governing_equations_disc}

\subsubsection{Equation of state}
\label{ssub:Governing_equations_EoS}

\newcommand{\cs}{c_{s}} 

The first two variables are the pressure $P$ and the surface density $\Sigma$. To keep things simple, we assume that the disc is globally isothermal of sound speed $\cs$, so that
\begin{equation}
    \label{eq:Equation_of_state}
    P = \cs^{2} \, \Sigma .
\end{equation}

\subsubsection{Equation of motion}
\label{ssub:Governing_equations_EoM}

\renewcommand{\ur}{u_{r}} 
\newcommand{\up}{u_{\varphi}} 

\newcommand{\phis}{\phi_{\star}} 
\newcommand{\phid}{\phi_{d}} 

The standard way to describe the disc is to use the cylindrical coordinates $r$ and $\varphi$, the surface density $\Sigma$, the radial velocity $\ur$, the azimuthal velocity $\up$, the central object's gravitational potential $\phis$ and the disc's self potential $\phid$. But we prefer another set of variables introduced in \cite{TaggerPellat99} that makes the equations lighter. It simply replaces $r$ by ${ s = \ln{(r / R_{0})} }$, $\ur$ by ${ U = r \, \ur }$, $\up$ by ${ V = r \, \up }$ and $\Sigma$ by ${ \sigma = \ln{(\Sigma / \Sigma_{0})} }$ where \smash{$R_{0}$} and \smash{${\Sigma_{0}}$} are reference radii and surface densities. The Eulerian mass and momentum conservation equations then write
\begin{subequations}
    \small
    \label{eq:EoM_disc}
    \begin{align}
        & \!\!\!\!\!\!\!\! \partial_{t} \sigma + \frac{1}{r^{2}} \big[ \partial_{s} U + U \partial_{s} \sigma + \partial_{\varphi} V +  V \partial_{\varphi} \sigma \big] = 0 , \label{eq:EoM_disc_mass} \\
        & \!\!\!\!\!\!\!\! \partial_{t} U \! + \! \frac{U}{r^{2}} \big[ \partial_{s} U \! - \! U \big] \! + \! \frac{V}{r^{2}} \big[ \partial_{\varphi} U \! - \! V \big] \! = \! - \cs^{2} \, \partial_{s} \sigma \! - \! \partial_{s} \big[ \phis \! + \! \phid \big] \! , \!\!\!\!\!\!\!\! \label{eq:EoM_disc_momentum_r} \\
        & \!\!\!\!\!\!\!\! \partial_{t} V \! + \! \frac{U}{r^{2}} \big[ \partial_{s} V \! - \! V] \! + \! \frac{V}{r^{2}} \big[ \partial_{\varphi} V \! + \! U \big] \! = \! - \cs^{2} \, \partial_{\varphi} \sigma \! - \! \partial_{\varphi} \big[ \phis \! + \! \phid \big] \! . \!\!\!\!\!\!\!\! \label{eq:EoM_disc_momentum_p}
    \end{align}
\end{subequations}

\subsection{Equations of motion for the central object}
\label{sub:Governing_equations_central_object}

We showed in our first paper that the choice of reference frame affects the number, stability, pattern frequency and growth rate of the reflex modes. We therefore need to introduce the equations governing the motion of the central object in both reference frames.

\subsubsection{Stellocentric reference frame}
\label{ssub:Governing_equations_central_object_stellocentric}

\newcommand{\phii}{\phi_{i}} 

\newcommand{\bas}{\ba_{\star}} 

In the stellocentric reference frame, the gravitational potential of the central object is constant,
\begin{equation}
    \label{eq:central_object_potential_stellocentric}
    \phis (\br) = - \frac{G \Ms}{r} ,
\end{equation}
but the reference frame is non-inertial, so we need to add a fictitious force to the disc's governing equations. This force is uniform, so it derives from the `indirect' potential
\begin{equation}
    \label{eq:indirect_potential}
    \phii = \bas \bcdot \br ,
\end{equation}
where $\bas$ is the acceleration that the central object would have if we were in the barycentric reference frame.

To measure this acceleration, we simply integrate the gravitational forces exerted by all the disc parcels,
\begin{equation}
    \label{eq:central_object_acceleration_stellocentric}
    \bas = \iint\limits_{\text{disc}} \! \rd \br \,\, \frac{G \, \Sigma(\br)}{r^{2}} \, \er ,
\end{equation}
where \smash{$\er$} is the radial unit vector.

\subsubsection{Barycentric reference frame}
\label{ssub:Governing_equations_central_object_barycentric}

\newcommand{\brs}{\br_{\star}} 

This reference frame is inertial, so there is no need for fictitious forces. The backlash, of course, is that the central object can move. It is only subject to the disc's gravity, so its equation of motion writes
\begin{equation}
    \label{eq:central_object_acceleration_barycentric}
    \rd_{tt}^{2} \, \brs = \iint\limits_{\text{disc}} \! \rd \br \,\,  \frac{G \, \Sigma(\br)}{||\br - \brs||^{2}} \, \frac{\br - \brs}{||\br - \brs||} ,
\end{equation}
where $\brs$ denotes the position of the central object and ${\rd \br}$ is a scalar surface element.

The gravitational potential of the central object is then 
\begin{equation}
    \label{eq:central_object_potential_barycentric}
    \phis (\br) = - \frac{G \Ms}{||\br - \brs||} .
\end{equation}

\subsection{Self-gravity}
\label{sub:Governing_equations_self_gravity}

In both reference frames, the self-gravitational potential of the disc is simply
\begin{equation}
    \label{eq:self_gravitational_potential}
    \phid (\br) = - \iint\limits_{\text{disc}} \! \rd \br' \,\, \frac{G \, \Sigma (\br')}{||\br' - \br||} .
\end{equation}

\section{Equilibrium state}
\label{sec:Equilibrium}

Our goal is to study the response of the system to tiny perturbations. But before we can do that, we need to define the equilibrium upon which infinitesimal perturbations might grow or propagate. We shall denote equilibrium variables with an overline ${ \overline{\phantom{a}} }$.

\newcommand{\oSigma}{\overline{\Sigma}} 
\newcommand{\our}{\overline{u}_{r}} 
\newcommand{\oup}{\overline{u}_{\varphi}} 

\newcommand{\os}{\overline{\sigma}} 
\newcommand{\oU}{\overline{U}} 
\newcommand{\oV}{\overline{V}} 

\newcommand{\obrs}{\overline{\br}_{\star}} 

In the absence of any perturbation, the central object is stationary (\smash{${\obrs = \mathbf{0} }$}) and the disc is circular (\smash{${\our = \mathbf{0}}$}). 

\newcommand{\ophid}{\overline{\phi}_{d}} 

The azimuthal velocity results from a balance between the central object's gravity, the centrifugal force, the radial pressure gradient and the disc's self-gravity, so we have \smash{${\oup = r \, \Omega(r) }$} with 
\begin{equation}
    \label{eq:rotation_frequency}
    \Omega (r) = \sqrt{\frac{G \Ms}{r^{3}} + \frac{\cs^{2}}{r} \frac{\rd_{r} \oSigma}{\oSigma} + \frac{1}{r} \rd_{r} \ophid} 
\end{equation}

\newcommand{\OmegaK}{\Omega_{\text{K}}} 

We control temperature via the dimensionless number \smash{${ h = \cs / [R_{0} \, \OmegaK (R_{0})] }$}, where \smash{${ \OmegaK (r) = \sqrt{G \Ms / r^{3}} }$} is the Keplerian frequency. Physically, $h$ represents the disc’s aspect ratio at \smash{${ r = R_{0} }$}.

\newcommand{\ri}{r_{\text{in}}} 
\newcommand{\ro}{r_{\text{out}}} 
\newcommand{\re}{r_{\text{in or out}}} 

Finally, we need to define the surface density profile. Following \cite{Crida+25b}, we assume that \smash{$\oSigma$} follows a power law of index ${p}$, \smash{${ \oSigma (r) \propto r^{-p} }$}, and ranging from ${ \ri}$ to ${\ro}$. The proportionality coefficient \smash{$\Sigma_{0}$} is set by the mass of the disc relative to the central object, ${ \Md / \Ms }$.

Unless stated otherwise, we use ${ p \! = \! 1.5 }$, ${ \ri / R_{0} \! = \! 0.32 }$, ${ \ro / R_{0} = 3.125 }$, ${ \Md / \Ms = 0.06 }$ and ${ h = 0.05 }$ in all our figures. In any case, the parameters we used for all the figures are listed in Appendix~\ref{sec:Parameters}.

\section{Linear perturbation equations}
\label{sec:Linear_perturbation_equations}

Let us now perturb the background flow, \smash{${ \overline{f} }$}, with a perturbation, \smash{${ f' }$}, which we assume to be small, \smash{${ f' \ll \overline{f} }$}. The total flow is then \smash{${ f = \overline{f} + f' }$}, where ${ f }$ denotes any variable (density, velocity, potential, \textit{etc.}).

\subsection{Equations of motion for the disc}
\label{sub:Linear_perturbation_equations_disc}

If we neglect the quadratic terms of the form \smash{${ f' \times f' }$}, the disc's governing equations~\eqref{eq:EoM_disc} become
\begin{small}
\begin{subequations}
    \label{eq:Linear_perturbation_equations}
    \begin{align}
        & \!\!\!\!\! \big( \partial_{t} + \Omega \, \partial_{\varphi} \big) \, \sigma' + \frac{1}{r^{2}} \left[ \partial_{s} U' + U' \, \rd_{s} \os + \partial_{\varphi} V' \right] = 0 , \label{eq:Linear_perturbation_equations_mass} \\
        & \!\!\!\!\! \big( \partial_{t} + \Omega \, \partial_{\varphi} \big) \, U' - 2 \Omega \, V' = - \cs^{2} \partial_{s} \sigma' - \partial_{s} \big[ \phis' + \phii' + \phid' \big] , \!\!\!\! \label{eq:Linear_perturbation_equations_momentum_r} \\
        & \!\!\!\!\! \big( \partial_{t} + \Omega \, \partial_{\varphi} \big) \, V' + \frac{\kappa^{2}}{2 \Omega} \, U' = - \cs^{2} \partial_{\varphi} \sigma' - \partial_{\varphi} \big[ \phis' + \phii' + \phid' \big] \! . \!\!\!\! \label{eq:Linear_perturbation_equations_momentum_p}
    \end{align}
\end{subequations}
\end{small}
\!\!\!\!where \smash{${ \kappa \! = \! \sqrt{ \rd_{r} [(r^{2} \Omega)^{2}] / r^{3} } }$} is the local epicyclic frequency.

\newcommand{\hSigma}{\hat{\Sigma}} 
\newcommand{\hur}{\hat{u}_{r}} 
\newcommand{\hup}{\hat{u}_{\varphi}} 

\newcommand{\hs}{\hat{\sigma}} 
\newcommand{\hU}{\hat{U}} 
\newcommand{\hV}{\hat{V}} 

\newcommand{\hphis}{\hat{\phi}_{\star}} 
\newcommand{\hphii}{\hat{\phi}_{i}} 
\newcommand{\hphid}{\hat{\phi}_{d}} 

\newcommand{\C}{\mathcal{C}} 

Those equations are linear and none of the coefficients depend on time or azimuth, so we can look for Fourier-mode solutions of the form \smash{${ f(t, r, \varphi) = \widehat{f} (r) \, \e^{i (\varpi t - m \varphi)} }$}. Equations~\eqref{eq:Linear_perturbation_equations} simplify to
\begin{subequations}
    \label{eq:Fourier-transformed_equations}
    \begin{align}
        & i \C \, \hs + \frac{1}{r^{2}} \left[ \rd_{s} \hU + \hU \, \rd_{s} \os - i m \, \hV \right] = 0 , \label{eq:Fourier-transformed_equations_mass} \\
        & i \C \, \hU - 2 \Omega \, \hV = - \cs^{2} \, \rd_{s} \hs - \rd_{s} \big[ \hphis + \hphii + \hphid \big] , \label{eq:Fourier-transformed_equations_momentum_r} \\
        & i \C \, \hV + \frac{\kappa^{2}}{2 \Omega} \, \hU = i m \, \cs^{2} \, \hs + i m \, \big[ \hphis + \hphii + \hphid \big] . \label{eq:Fourier-transformed_equations_momentum_p}
    \end{align}
\end{subequations}
where ${ \C = \varpi - m \Omega }$ is the Doppler-shifted frequency. We use the letter $\C$ to indicate that it is zero at corotation.

\subsection{Equations of motion for the central object}
\label{sub:Linear_perturbation_equations_central_object}

\subsubsection{Stellocentric reference frame}
\label{ssub:linear_perturbation_equations_central_object_stellocentric}

\newcommand{\hbas}{\hat{\ba}_{\star}} 

Since linear analysis of the central object's dynamics is uncommon, let us go through it slowly. We start from Eq.~\eqref{eq:central_object_acceleration_stellocentric}, whose only linear term is
\begin{equation}
    \nonumber
    \bas' = \iint\limits_{\text{disc}} \! \rd \br \,\, \frac{G \, \Sigma'(\br)}{r^{2}} \, \er .
\end{equation}

We then replace \smash{$\er$} by \smash{${ \cos{(\varphi)} \, \ex + \sin{(\varphi)} \, \ey }$}, \smash{$\Sigma'$} by \smash{${ \hSigma (r) \, \e^{i (\varpi t - m \varphi)} }$} and \smash{${ \bas' }$} by \smash{${ \hbas \, \e^{i \varpi t} }$} to get
\begin{subequations}
    \begin{flalign}
    & \!\!\!\!\!\!\!\!\! \hbas = 
    G \int_{\ri}^{\ro} \!\!\!\!\! r \rd r \, \frac{\hSigma (r)}{r^{2}} \int_{0}^{2 \pi} \!\!\! \rd \varphi \, \big[\cos{(m \varphi)} - i \, \sin{(m \varphi)} \big] \nonumber \\
    \pushright{\displaystyle \times \big[ \cos{(\varphi)} \, \ex + \sin{(\varphi)} \, \ey \big] .} \nonumber \\
    & \!\!\!\!\!\!\!\!\! \phantom{\hbas} = 
        \begin{cases}
            \displaystyle \pi G \, \int_{\ri}^{\ro} \!\!\!\!\! \rd r \, \frac{\hSigma (r)}{r} \left[\ex \mp i \, \ey \right] & \text{ if } m = \pm 1 , \\
            0 & \text{ otherwise} .
        \end{cases} \nonumber
    \end{flalign}
\end{subequations}
This shows that only the \smash{${ m = 1 }$} component of disc perturbations can cause a reflex motion of the central object.

Finally, we inject this into Eq.~\eqref{eq:indirect_potential} and replace \smash{${ \br }$} by \smash{${ r \left[ \cos{(\varphi)} \, \ex + \sin{(\varphi)} \, \ey \right] }$} to get \addtocounter{equation}{-1}
\begin{small}
\begin{equation}
    \label{eq:Fourier_transformed_equations_indirect_potential}
    \hphii (r) = \delta_{m}^{\pm 1} \, \pi G r \int_{\ri}^{\ro} \!\!\!\!\! \rd r' \, \frac{\hSigma (r')}{r'} .
\end{equation}
\end{small}

\newcommand{\hIT}{\hat{\text{IT}}} 

Instead of the indirect potential, we will sometimes use the `indirect term'
\begin{small}
\begin{equation}
    \label{eq:Fourier_transformed_equations_indirect_term}
    \hIT = \frac{- \hphii (r) \, / \, r}{R_{0} \, \OmegaK^{2} (R_{0}) }
\end{equation}
\end{small}
\!\!\!\!where ${ \OmegaK }$ is the Keplerian frequency. \smash{$\hIT$} has two advantages: it is independent of $r$, and it is dimensionless.

\subsubsection{Barycentric reference frame}
\label{ssub:linear_perturbation_equations_central_object_barycentric}

\newcommand{\brsi}{\brs'} 
\newcommand{\hbrs}{\hat{\br}_{\star}} 

Following similar steps in the other reference frame, we write \smash{${ \brsi (t) = \hbrs \, \e^{i \varpi t} }$} and reduce Eq.~\eqref{eq:central_object_acceleration_barycentric} to
\begin{equation}
    \label{eq:Fourier_transformed_equations_central_object_location_barycentric}
    \hbrs = - \frac{\delta_{m}^{\pm 1} \, \pi G}{\varpi^{2} + \mu^{2}} \int_{\ri}^{\ro} \! \rd r \, \frac{\hSigma}{r} \, \left[ \ex \mp i \ey \right]
\end{equation}
and Eq.~\eqref{eq:central_object_potential_barycentric} to
\begin{equation}
    \label{eq:Fourier_transformed_equations_central_object_potential_barycentric}
    \hphis (r) = \frac{\delta_{m}^{\pm 1} \, \pi G^{2} \Ms}{r^{2} \left(\varpi^{2} + \mu^{2}\right)} \int_{\ri}^{\ro} \!\!\!\! \rd r' \, \frac{\hSigma (r')}{r'} .
\end{equation}
The full derivation can be found in our first paper. \smash{${ \mu = \scriptstyle \sqrt{ \pi G \int \! \rd r \, \oSigma / r^{2}} }$} is a frequency indicative of how unstable an equilibrium the unperturbed disc centre is.

\subsection{Self-gravity}
\label{sub:Linear_perturbation_equations_self_gravity}

Since the surface density perturbation is of the form \smash{${ \hSigma (r) \, \e^{i (\varpi t - m \varphi)} }$}, we can simplify Eq.~\eqref{eq:self_gravitational_potential} to
\begin{equation}
    \label{eq:Linear_perturbation_equations_self_gravity_potential}
    \hphid (r) = - G \int_{\ri}^{\ro} \rd r' \, K_{m} \left( \frac{r'}{r} \right) \, \hSigma (r') ,
\end{equation}
where
\begin{equation}
    \label{eq:Gravitational_kernel}
    K_{m} (x) = x \int_{0}^{2 \pi} \rd \varphi \,\, \frac{\cos{(m \varphi)}}{\sqrt{1 + x^{2} - 2 x \cos{(\varphi)}}}
\end{equation}
is the gravitational kernel. It has a singularity in~${ x = 1 }$, so we regularise it by adding an~$\epsilon^{2}$ term inside the square root, with ${ \epsilon = 10^{-2} }$.

\vspace{-0.33 \baselineskip}
\subsection{Boundary conditions}
\label{sub:linear_perturbation_equations_BCs}

Because of pressure, 2D discs support density waves that make the linear problem hyperbolic and create a need for \BCs. These may be imposed in \smash{${\ri}$} or in \smash{${\ro}$}, and may affect \smash{$\hs$}, \smash{$\hU$} or \smash{$\hV$}. We restrict ourselves to Dirichlet boundary conditions (\smash{${ \hat{f} = 0 }$}) and Neumann boundary conditions (\smash{${ \rd_{s} \hat{f} = 0 }$}), except in Fig.~\ref{fig:RWI} where we use a purpose-built radiative \BC.

\vspace{-0.33 \baselineskip}
\subsection{Damping layers}
\label{sub:linear_perturbation_equations_damping_layers}

The density waves can sometimes reflect off the boundaries. To avoid this, simulations frequently use damping layers. The idea is to add a dissipative term \smash{${ - D(r) \hat{f} }$} to the right-hand-side of the disc's linear perturbation equations~\eqref{eq:Fourier-transformed_equations}, in order to damp all waves before they can reflect. Unfortunately, this is tricky because the damping profile ${ D(r) }$ must have a large maximum (so that all waves die), be null everywhere except close to the boundary (so that most of the domain follows the true equations), and have no large derivative (so that waves do not reflect off the boundary layer itself).

\newcommand{\rid}{r_{\text{in}}^{\text{d}}} 
\newcommand{\rod}{r_{\text{out}}^{\text{d}}} 
\newcommand{\red}{r_{\text{in or out}}^{\text{d}}} 

Our damping layers act on all three variables~\smash{$\hSigma$}, \smash{$\hU$} and \smash{$\hV$}. The inner damping layer ranges from~\smash{$\ri$} to~\smash{${ \rid = 0.4 R_{0} }$}, and the outer damping layer from \smash{${ \rod = 2.5 R_{0} }$} to~\smash{$\ro$}. The damping profiles follow
\begin{equation}
    \label{eq:linear_perturbation_equations_damping}
    D(r) = d \times \left( \frac{\red - r}{\red - \re} \right)^{2} \times \OmegaK (r) .
\end{equation}
The damping level~$d$ is nominally ${ 10 / 3 },$ but we can set it to zero to deactivate the damping layers.

\vspace{-0.33 \baselineskip}
\section{Exploration strategy}
\label{sec:Exploration_strategy}

\begin{figure}
    \centering
    \includegraphics[width = \linewidth]{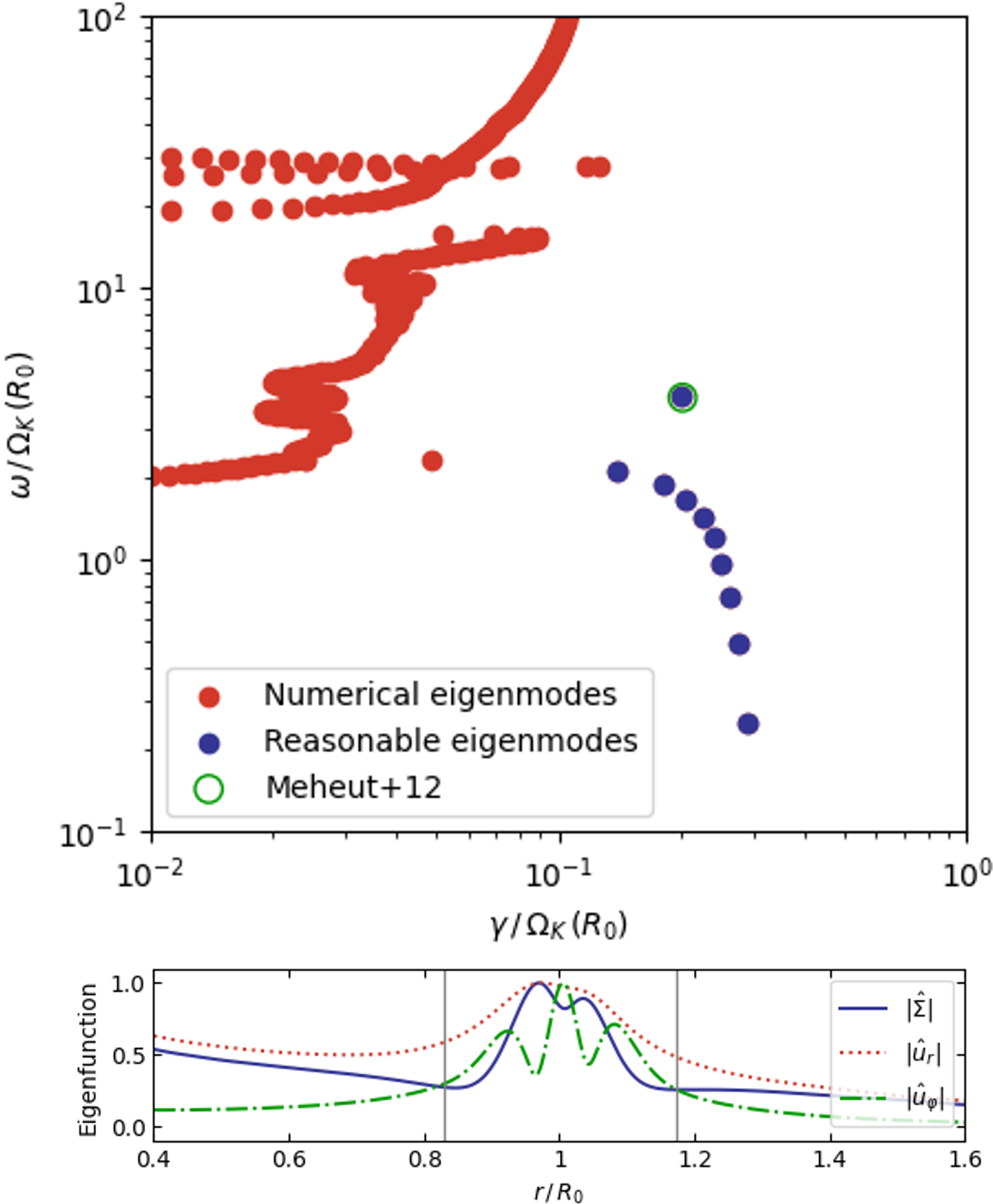}
    \caption{
        \textit{Top:} Eigenvalues found by our new linear analysis code LAADS in the \smash{${ m = 4 }$} \RWI\ setup of \cite{Meheut+12}. Like all linear analysis code, LAADS solves a discretised form of the linear perturbation equations, so the number of eigenmodes it finds is directly given by the resolution of the discretisation. We do not know how many eigenmodes the continuous linear perturbation equations admit, but usually a lot less. Distinguishing between discretisation artefacts and physical modes is a hard problem which, to the best of our knowledge, has not been solved yet. LAADS runs a few sanity checks to eliminate the bulk of the artificial modes (red dots), but we cannot guarantee that there is no false negative or false positive. In this particular case, it seems to work well: there is a sequence of nine physical-looking modes and a single isolated \RWI\ mode whose eigenvalue falls within 1\% of Meheut's. However, as we explore the high-dimensional parameter space of reflex instabilities, it would be extremely time-consuming to investigate the physics of all the reasonable modes that appear. For instance here we did not examine the nature of the sequential modes.
        \textit{Bottom:} Partial reproduction of figure~3 in \cite{Meheut+12}. It shows the eigenfunction of the \RWI\ mode from the top panel. The thin grey lines give the positions of the Lindblad resonances. Far from the boundaries, LAADS' and Meheut's eigenfunctions are indistinguishable. Near the boundaries they differ slightly because Meheut's calculations are 3D while ours are 2D.
    }
    \label{fig:RWI}
    \vspace{-1 \baselineskip}
\end{figure}

To power our exploration of parameter space, we use a new code called LAADS for \textit{Linear Analysis of Astrophysical DiscS} and which solves Eqs.~\eqref{eq:Fourier-transformed_equations}. We will present this code in a future publication, but since it has not gone through the peer-review process yet, our reader may question its correctness. In order to build trust, we invite our reader to look at figure~1 in our first paper. It shows that our code is in excellent agreement with our analytical description of the `bracket' modes. We also invite our reader to read Appendix~\ref{sec:Analytical_description_of_the_cavity_modes} of the present paper, where we derive an analytical description of the `cavity' modes and show that once again, our code is in good agreement with the analytics. Finally, Fig.~\ref{fig:RWI} shows that LAADS can quantitatively reproduce the eigenvalues and eigenvectors of known accretion disc instabilities such as the \RWI.

Equations~\eqref{eq:Fourier-transformed_equations} contain five numerical parameters, one function parameter, and sixteen boolean parameters. Furthermore, for a given set of parameters, we do not know \textit{a priori} how many modes the linear perturbation equations~\eqref{eq:Fourier-transformed_equations} will admit. All of this makes it impossible to explore parameter space exhaustively. We shall instead focus on the modes that appear in the fiducial disc of \S\ref{sec:Equilibrium}, plus the reflex instabilities that are already known thanks to \cite{Adams+89}, \cite{Heemskerk+92}, \cite{Sanchez+25} and \cite{Crida+25b}.

We present our findings by order of increasing complexity: paper one neglected pressure, self-gravity and the damping layers. We first reactivate pressure in \S\ref{sec:Bracket_modes} and \S\ref{sec:Cavity_modes}, then self-gravity in \S\ref{sec:Gravitational_modes}, and finally the damping layers in \S\ref{sec:Overdamped_modes}. We shall see that new instabilities appear each time we increase complexity.

\section{Bracket modes}
\label{sec:Bracket_modes}

We already studied the pressure-less version of bracket modes in our first paper. They are exact solutions to the pressure-less version of Eqs.~\eqref{eq:Fourier-transformed_equations}, but they are  also artefacts caused by the boundaries of the domain being inside the disc. Indeed, since mass can easily leave the domain, it is not conserved. The bracket mode of the barycentric equations is unstable, so it can quickly cause a lot of damage to simulations. The stellocentric bracket mode is marginally stable, but can still act as an artificial store of energy that steals some of the energy that would otherwise be released into the turbulent cascade by non-linear interactions between genuine modes. Therefore, the amplitude of some of the modes that take part in the cascade may be misestimated. Unfortunately, those artefacts are nearly unavoidable, especially in discs that lack a central cavity.

The present section focuses on the pressurised version of the modes. The key difference compared to the pressure-less case is that we must impose \BCs. Unfortunately, we showed in paper one (Eqs. 18, 19, 21 and 22) that \smash{$\hat{f}$} and \smash{${\rd_{r} \hat{f}}$} are both non-zero at the boundaries, for all variables \smash{${ \hat{f} \in \{\hs, \, \hU, \, \hV \} }$}. That makes it difficult to connect the pressure-less and pressurised cases. 

\begin{figure*}
    \centering
    \includegraphics[width = \linewidth]{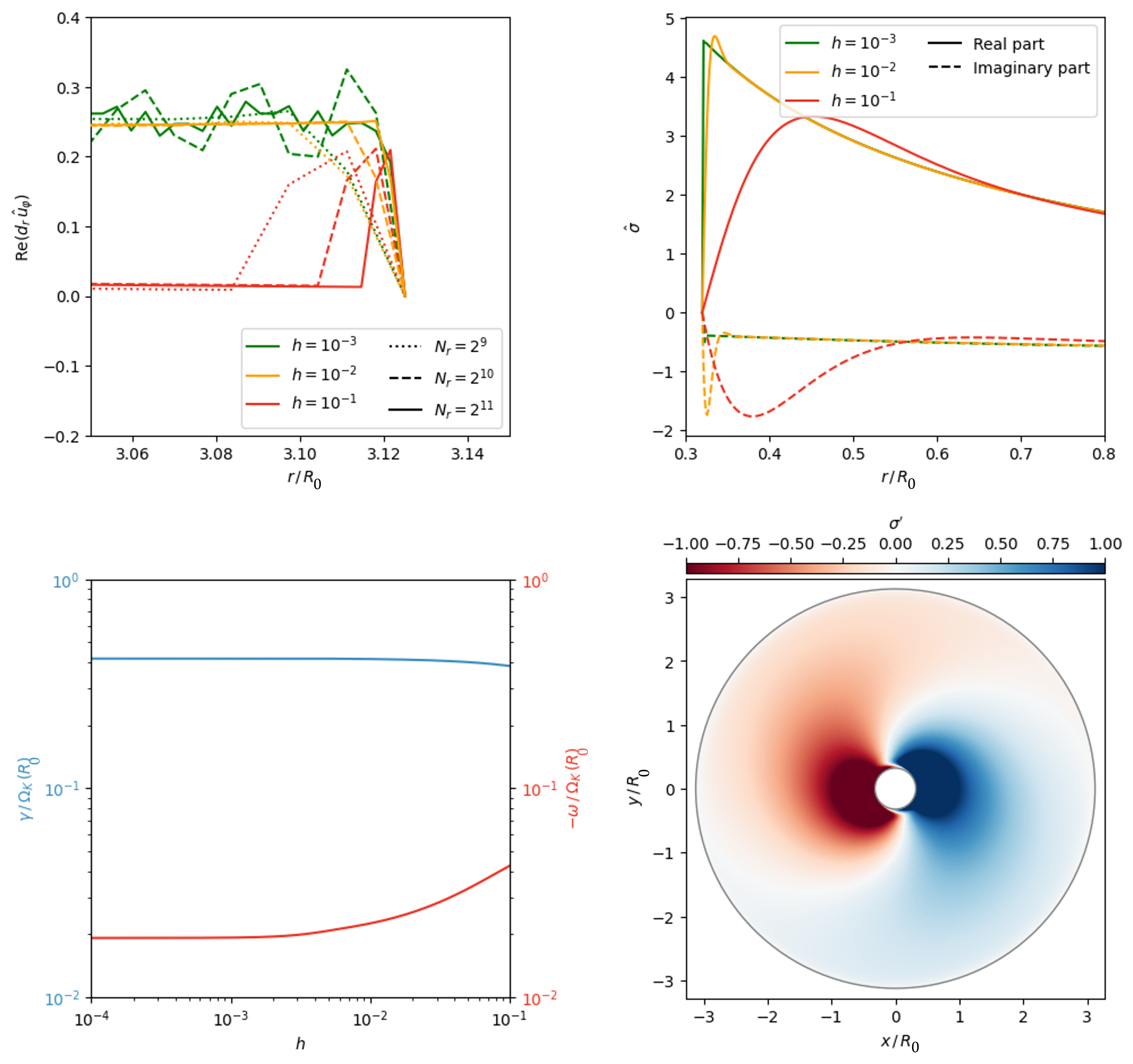}
    \caption{
        \textit{Top left:} Radial derivative of the azimuthal velocity perturbation \smash{${ \rd_{r} \hup }$} induced by our code's tentative pressurised stellocentric bracket mode. We choose this observable because we imposed Neumann boundary conditions on \smash{${ \hup }$}. We vary the temperature of the disc via $h$ and the resolution of the solver via $N_{r}$. We see concerning features in all cases: oscillations at low temperature, a discontinuity at intermediate temperature, and a bump at high temperature. None of those features converges with resolution.
        \textit{Top right:} Same, but for the relative density perturbation $\hs$ induced by the unstable barycentric bracket mode under Dirichlet \BCs\ on $\hs$. This time, the eigenfunction looks smooth at all temperatures.
        \textit{Bottom left:} Growth rate \smash{${ \gamma = - \Im (\varpi) }$} and pattern frequency \smash{${ \omega = \Re (\varpi) }$} of the unstable barycentric bracket mode as a function of temperature. Note that the growth rate is the same for all \BCs\ except \smash{${ \hur = 0 }$}. The pattern frequency, on the other hand, is very sensitive to the \BCs.
        \textit{Bottom right:} 2D representation of the relative density perturbation induced by the unstable barycentric bracket mode in our fiducial (pressurised) disc. Remember that the maximum amplitude of the eigenvector is a gauge choice, it does not have any physical meaning. What matter are the relative amplitudes between the perturbations to different variables and at different radii. The parameters used to produce each plot in this and all subsequent figures are listed in Appendix~\ref{sec:Parameters}.
    }
    \label{fig:Bracket_modes}
\end{figure*}

In fact, we could not find convincing evidence for pressurised bracket modes in the stellocentric reference frame. We used a continuation algorithm to find the pressurised eigenvalue that connects to the pressure-less eigenvalue, but Fig.~\ref{fig:Bracket_modes}-\textit{top left} shows that the corresponding pressurised eigenvector has suspicious features such as rapid oscillations and jumps, and no sign of convergence with resolution. This suggests that this pressurised eigenmode is a numerical artefact induced by the discretisation of the eigenvalue problem formed by Eqs.~\eqref{eq:Fourier-transformed_equations}~and~\eqref{eq:Fourier_transformed_equations_indirect_potential}. We therefore conjecture that the stellocentric reference frame does not allow for pressurised bracket modes.

The barycentric bracket mode, however, looks reasonable even at low resolution. Fig.~\ref{fig:Bracket_modes}-\textit{top right} shows that its pressurised eigenfunction operates a smooth connection between the pressure-less solution and the boundary condition. Fig.~\ref{fig:Bracket_modes}-\textit{bottom left} shows that temperature has little effect on the growth rate and on the pattern frequency. Finally, Fig.~\ref{fig:Bracket_modes}-\textit{bottom right} displays the relative density perturbation $\sigma'$ induced by the unstable pressurised barycentric bracket mode in our fiducial disc. It has a loosely-wound, ${m = 1 }$ spiral structure.

\section{Cavity modes}
\label{sec:Cavity_modes}

Density waves can propagate in the disc and reflect at the boundaries. Repeated reflection turns the disc into a cavity, in direct analogy with the standing light waves of a laser cavity. And since the reflex motion of the central object continuously injects energy into this cavity, instabilities can develop. We start our analysis with this energy injection process.

\cite{Ostriker+92} used the method of stationary phase to show that the indirect term preferentially excites density waves at the Lindblad resonances (\textit{i.e.} where \smash{${ m \, \Omega (r) \pm \kappa (r) = \varpi }$}) and at the boundaries of the domain. Now accretion discs are thin, so unless a substructure creates a large (local) gradient, \smash{$\Omega$} and \smash{$\kappa$} are nearly equal everywhere. One consequence of this is that low-frequency ${ m = 1 }$ modes are close to the inner Lindblad resonance everywhere. Therefore, to study how accretion discs respond to slowly-varying indirect terms,\footnote{or, if we are in the barycentric reference frame, to slow movements of the central object} we need to solve a wave-propagation problem where forcing is efficient everywhere. 

We formulate this mathematical problem in Appendix~\ref{sec:Analytical_description_of_the_cavity_modes}. It takes the form of a linear second-order \ODE\ with varying coefficients and non-localised forcing (Eq.~\ref{eq:Cavity_modes_ODE}). We use WKB analysis \citep{BenderOrszag99} to construct an approximate analytical solution. Due to the delocalised forcing, the particular solution is a complex mix of density waves of all phases. Fortunately, we can approximate it to a high degree of precision by the sum of the pressure-less response plus two density waves of equal amplitude propagating in opposite direction (Eq.~\ref{eq:WKB_particular_solution}). The two homogeneous solutions are also density waves propagating in opposite directions (Eq.~\ref{eq:WKB_solution_homogneous}), but their amplitude can vary depending on the boundary conditions.

As we explored parameter space, we found three types of cavity modes. We shall briefly describe each of them in a dedicated subsection.

\subsection{Resonant cavity modes}
\label{sub:Cavity_modes_resonant}

In Appendix~\hyperref[sub:Analytical_description_of_the_cavity_modes_resonant]{A.3}, we show that there exists a discrete set of (real) frequencies for which the boundary conditions cannot be satisfied. We use the letter $n$ to index them: \smash{$\omega_{1}$}, \smash{$\omega_{2}$}, ..., \smash{$\omega_{n}$}, ...

\begin{figure*}
    \centering
    \includegraphics[width = 0.81 \linewidth]{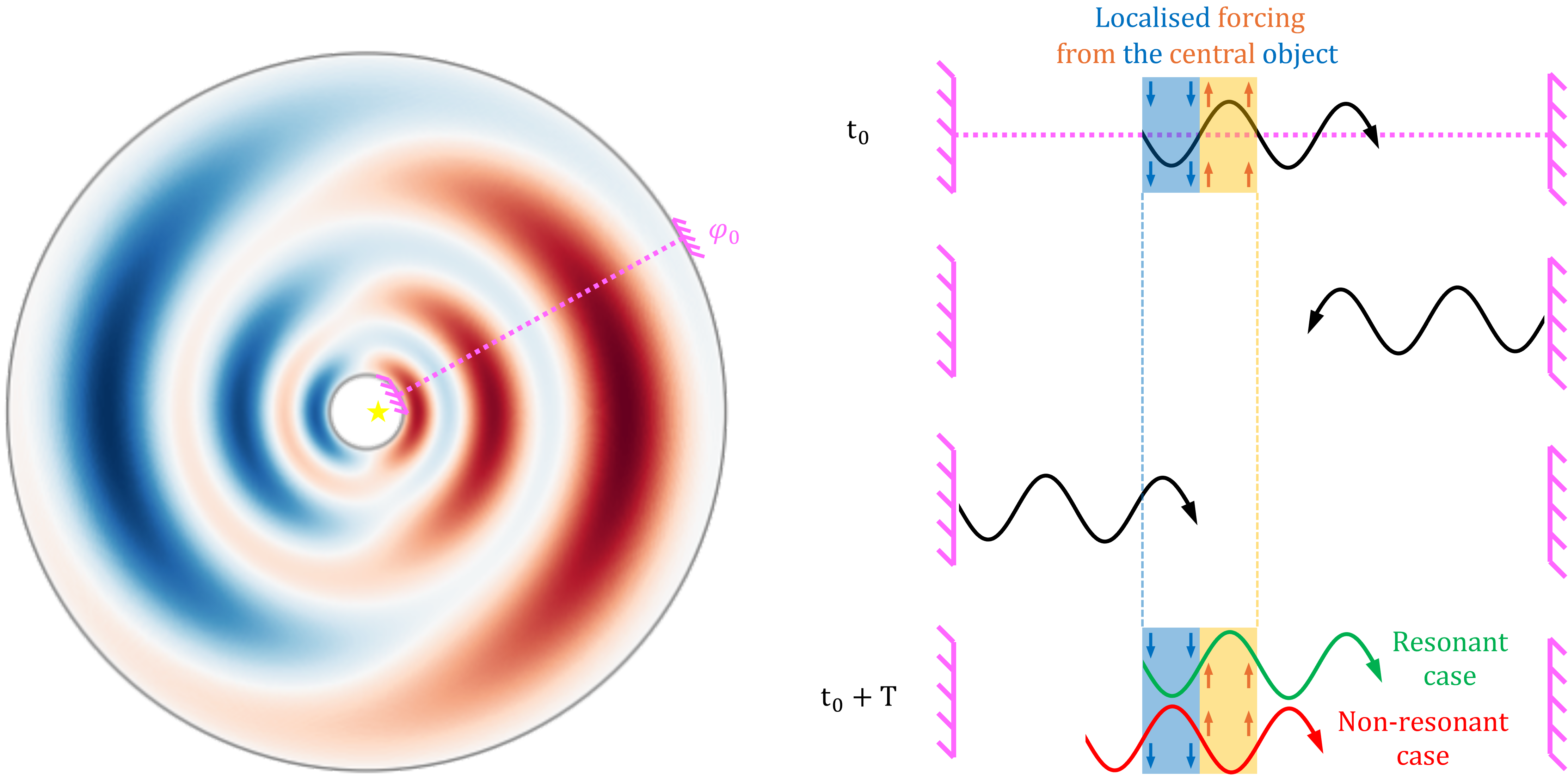}
    \caption{
        Diagram explaining why some cavity modes are resonant. We focus on what happens along a line of arbitrary azimuth \smash{${ \varphi_{0} }$} over time. At an arbitrary time \smash{$t_{0}$}, the forcing from the central object has the right spatial structure to emit a train of density waves propagating outward. This train will reflect on the outer boundary, then on the inner boundary, and eventually come back to its original location. Now the central object is moving periodically, so at time \smash{${ t_{0} + T }$} it exerts exactly the same forcing as at time \smash{${ t_{0} }$}. If the train comes back to its original location at that exact time (green case), the forcing makes it even stronger. Since this process repeats again and again, we get an algebraic instability. Conversely, if the train takes a little bit more than \smash{${ T }$} to come back to its original location (red case), then the forcing makes it weaker. After some time, the train converges to an equilibrium amplitude and phase-shift relative to the forcing. The closer we are to the resonant frequency, the larger the amplitude and the smaller the phase-shift. Of course, this picture is extremely simplified: the density waves should be global, the forcing should not be localised, and wave-forcing interactions happen at all times, not just at~\smash{${ t_{0} }$} and~\smash{${ t_{0} + T }$}. Nevertheless, our diagram gives the right intuition: the amplitude of the density wave is usually controlled by phase-mixing, but this mechanism fails in the resonant case.
        }
    \label{fig:Resonance}
    \vspace{-0.5 \baselineskip}
\end{figure*}

Figure~\ref{fig:Resonance} explains why. Essentially, if we forget for a moment that the disc has an effect on the central object's motion, we can imagine that the central object describes a circle and exerts a periodic force on the disc, which responds by emitting density waves. At a given instant, this forcing injects energy into the wave that has a specific phase. Usually, this phase varies over time without favouring any particular wave. Since the waves of phase \smash{${ \phi_{0} }$} and \smash{${ \phi_{0} + \pi }$} are as likely as each other to receive energy, phase mixing prevents secular growth. But in the resonant case the forcing always injects energy into the same wave, which can then grow without limit. Therefore, we expect the disc's response to grow linearly in time. This is called an algebraic instability.

Now if the disc's response grows linearly with time, it is not a Fourier mode. This explains why we cannot find a WKB solution that satisfies the boundary conditions. Indeed, our WKB analysis only explores a subset of the Fourier solutions.

What might save us is that the algebraically growing mode in the disc affects the movement of the central object. This additional feedback loop may affect the resonant modes. If it is positive, it might make them exponentially rather than algebraically unstable, in which case they return to Fourier space. This is the mechanism behind resonant drag instabilities \citep{Magnan+24}. 

We shall see in the next subsection that the collisional cavity modes are most unstable when their pattern frequency is close to the resonant frequency. We therefore suspect that they are the exponentially unstable form of the resonant modes. 

Unfortunately, we cannot prove this convincingly. Indeed, to investigate algebraic instabilities, one needs to use Laplace transforms instead of Fourier transforms. This is why we cannot conclusively affirm that resonant modes exist, nor that they are unstable. We can only highlight their potential.

\subsection{Collisional cavity modes}
\label{sub:Cavity_modes_collisional}

Retrograde marginally stable modes (\textit{i.e.} those with \smash{${ \omega < 0}$} and \smash{${ \gamma = 0}$}) are sometimes well-ordered. Indeed, as we decrease the pattern frequency, we find two modes, two resonance frequencies, two modes, two resonance frequencies, and so on. 

Now if we follow a pair of modes and increase the disc's temperature, we see them collide with each other (both in eigenvalue and in eigenfunction). If we increase the temperature even further we find that two modes emerge from the collision. They have the same pattern frequency as during the collision, but non-zero growth rates: one mode is unstable, the other one is damped. Furthermore, if we plot the residual of Eq.~\eqref{eq:WKB_loop_closing_equation} -- \textit{i.e.} how far a trial frequency is from closing the disc/central object feedback loop -- along the marginally stable line -- \textit{i.e.} \smash{${\gamma = 0}$} -- we get a real-valued function of \smash{$\omega$} whose zeros mark the eigenfrequencies of marginally stable modes. We find that between two resonant frequencies~\smash{$\omega_{n}$} and~\smash{$\omega_{n+1}$}, the residual takes a quasi-quadratic form. This quadratic curve moves up as we increase the temperature, so the residual has two distinct real roots at low temperatures, a double real root at the collision temperature, and no real root at high temperature. All of those observations remind us of Krein collisions \citep{Krein50}, hence why we call those modes `collisional'. 

Surprisingly, if we keep increasing the temperature and follow the unstable and damped modes, we find that their growth rate reaches a maximum when their pattern frequency becomes resonant. After this point, the two modes start to converge towards the marginally stable line until a second Krein collision makes them marginally stable again. If they were originally marginally stable between the $n$-th and ${n+1}$-th resonances, they are now marginally stable between the ${n-1}$-th and $n$-th resonances. And the process repeats itself: as we keep increasing the temperature, the pair of modes can collide with each other over and over again. However, every time they go down the resonance ladder, they lose half a radial oscillation.

\begin{figure*}
    \centering
    \includegraphics[width = \linewidth]{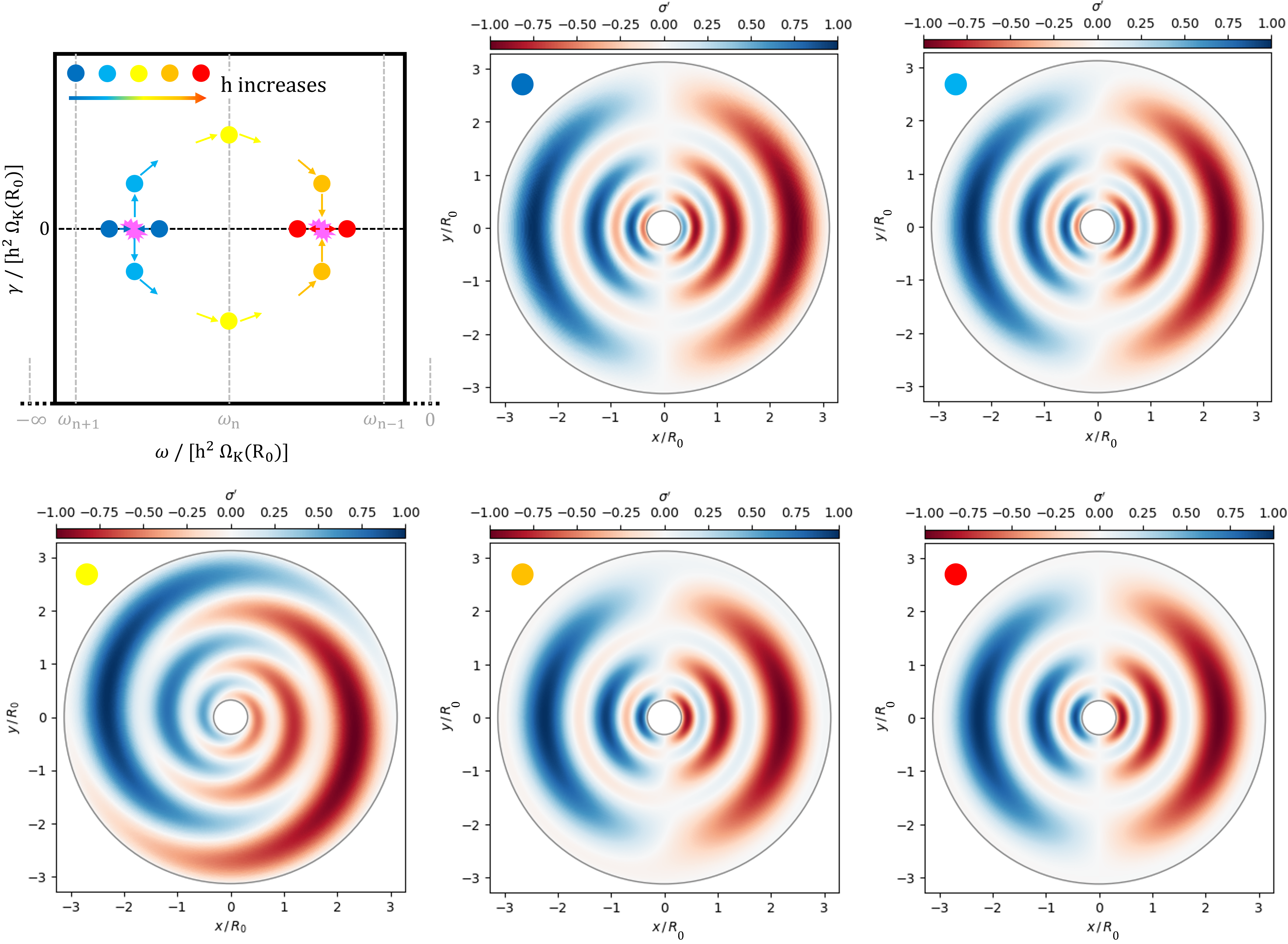}
    \caption{
        \textit{Top left:} Diagram explaining the evolution of a pair of collisional modes as the disc temperature increases (from dark blue to bright red). The marginally stable dark blue modes are both located between resonances $n$ and ${n + 1}$. They collide with each other to form the unstable and damped light blue modes. Those modes stray further and further from the marginally stable line while their (normalised) pattern frequency increases. Once it is close to resonance, we get the yellow modes, which are the most unstable/damped. At even higher temperatures, the orange modes start to converge back towards the marginally stable line. They eventually collide to form the marginally stable bright red modes. This process could repeat itself several times in the sense that at high enough temperature, the bright red modes could collide with each other to go further down the resonance ladder.
        \textit{Other panels:} 2D representation of the relative density perturbations induced by collisional modes representative of the five evolutionary stages presented in the top left panel. For the dark blue and bright red modes, we choose the highest-pattern-frequency mode of the pair. For the light blue, yellow and orange modes, we choose the unstable member of the pair. Notice that the bright red mode has one less half-radial-oscillation than the dark blue one (\smash{${ n = 6 \rightarrow 5 }$}).
        }
    \label{fig:Collisional_modes}
\end{figure*}

This whole process is depicted visually in Fig.~\ref{fig:Collisional_modes}-\textit{top left}, along with the eigenmodes associated with the different stages. It appears that the marginally stable modes create large-scale crescents, while the unstable modes create spirals. The number of crescents and the tightness of the spirals is directly linked to $n$.

\subsection{Standard cavity modes}
\label{sub:Cavity_modes_standard}

\begin{figure*}
    \centering
    \includegraphics[width = \linewidth]{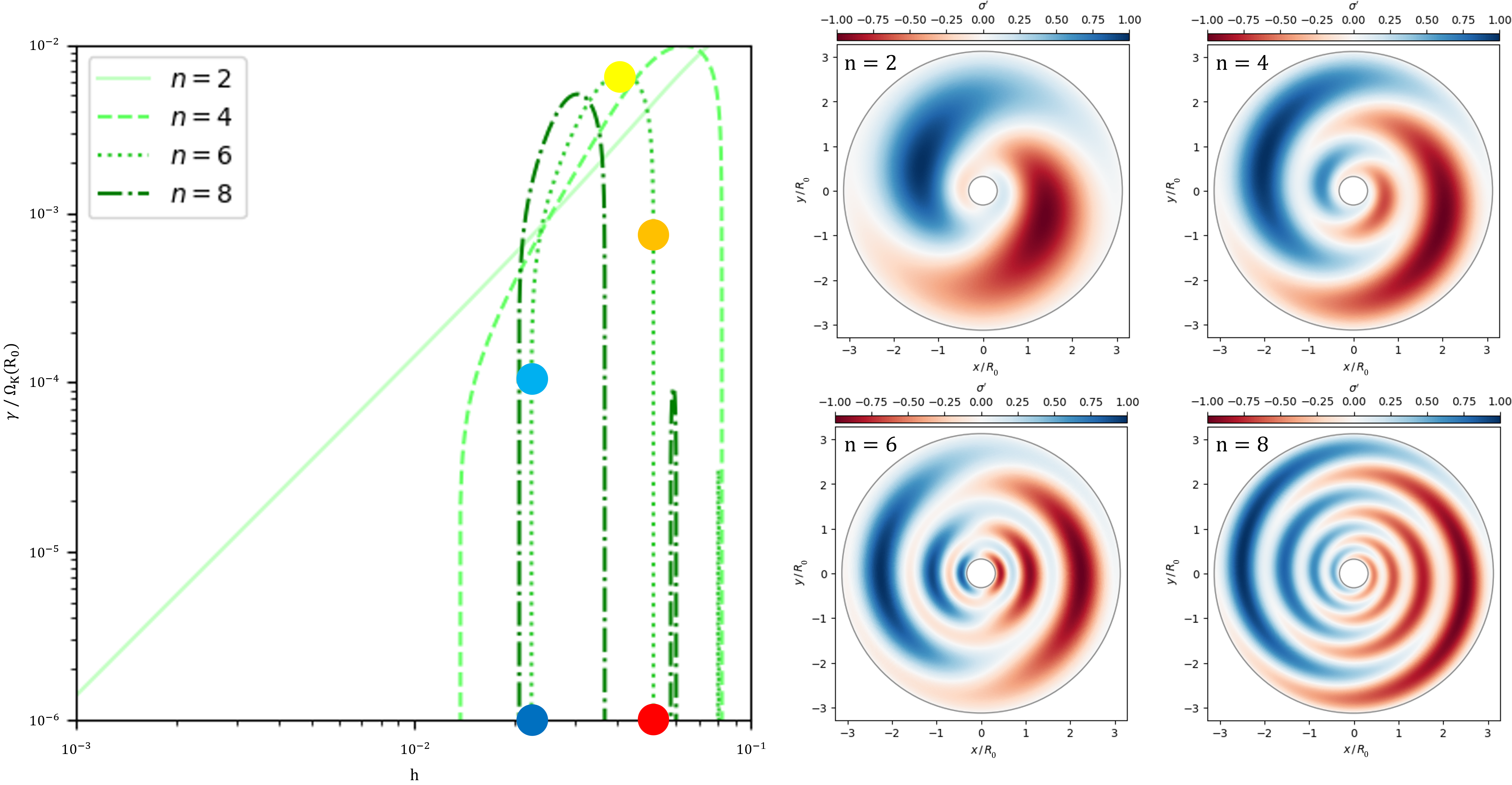}
    \caption{
        \textit{Left:} Growth rate $\gamma$ of the first four unstable stellocentric cavity modes as a function of disc temperature. We can see that the ${n = 4, \, 6, \, \text{and} \,\, 8}$ modes become unstable at a certain temperature and return to marginal stability at a higher temperature, as expected of collisional modes. Interestingly, we can see a second cycle at high temperature for the ${n = 6}$ and ${n = 8}$ modes, as claimed in \S\ref{sub:Cavity_modes_collisional}. We overlaid coloured circles to the first cycle of the ${n = 6}$ mode to facilitate the link to Fig.~\ref{fig:Collisional_modes}. Finally, the ${n = 2}$ mode is always unstable, so it is probably not a collisional mode.
        \textit{Other panels:} 2D representation of the relative density perturbations induced by the four modes of the left panel. The ${n = 2, \, 4, \, \text{and} \, 6}$ were measured at ${h = 0.05}$, the ${n = 8}$ mode at ${h = 0.03}$.
        }
    \label{fig:Cavity_modes}
    \vspace{-0.5 \baselineskip}
\end{figure*}

We plot in Fig.~\ref{fig:Cavity_modes}-\textit{left} the growth rates of the ${ n = 2, \, 4, \, 6 \,\, \text{and} \,\, 8}$ cavity modes as a function of temperature. The ${ n = 4, \, 6 \,\, \text{and} \,\, 8}$ modes are marginally stable below a certain temperature and above another temperature. We can therefore safely classify them as collisional modes. The ${n = 2}$ mode, however, is always unstable with a growth rate that scales with $h^{2}$. We pushed our exploration all the way down to \smash{${ h = 10^{-6} }$} (where LAADS hits its machine precision limit) but could not find any hint of a collision in this mode's low-temperature history. We therefore conjecture that it is not a collisional mode.

The core of its mechanism is probably the same as for every cavity mode: the displaced central object exerts a perturbative force on the disc, the disc responds by emitting density waves, the waves induce density perturbations, and these over/underdensities pull gravitationally on the central object and displace it even further from the origin. 

What we are unclear about are the details. On the one hand, the ${n = 2}$  mode does not seem to interact with either the ${n = 2}$ or the ${n = 3}$ resonances. On the other hand, the panels on the right of Fig.~\ref{fig:Cavity_modes} show a striking similarity between the azimuths selected by this mode and the collisional ones. The first observation suggests a difference between the physics of the collisional and standard modes, the second suggests a correspondence.

\section{Self-gravity-induced modes}
\label{sec:Gravitational_modes}

\cite{Crida+25a} made a compelling argument that whenever one accounts for the reflex motion of the central object in their model, they should also account for the disc's self-gravity. This activates (at least) two classes of prograde unstable modes. The SLING instability was discovered by~\cite{Adams+89} and the `evanescent' instability by ~\cite{Heemskerk+92}. We shall describe them in~\S\ref{sub:Gravitational_modes_SLING} and \S\ref{sub:Gravitational_modes_evanescent} respectively. 

Both rely on the $Q$-barrier, a radial area contained between the inner and outer Lindblad radii where density waves cannot propagate.

\subsection{SLING modes}
\label{sub:Gravitational_modes_SLING}

Following \cite{Adams+89}'s discovery, \cite{Shu+90} proposed a tentative explanation of the SLING instability's mechanism. Essentially, it uses the outer disc (\textit{i.e.} beyond corotation) as a resonant cavity: 
\begin{itemize}
    \vspace{-0.5 \baselineskip}
    \item[(i)] the cycle starts with an inward-propagating long density wave at the outer Lindblad resonance,
    \vspace{-0.5 \baselineskip}
    \item[(ii)] this long wave reaches the $Q$-barrier and refracts into an outward-propagating short density wave,
    \vspace{-0.5 \baselineskip}
    \item[(iii)] this short wave reaches the domain's edge and refracts into an inward-propagating short wave,
    \vspace{-0.5 \baselineskip}
    \item[(iv)] this short wave reaches the $Q$-barrier and refracts into an outward-propagating long wave,
    \vspace{-0.5 \baselineskip}
    \item[(v)] this wave reaches the outer Lindblad resonance and reflects into an inward-propagating long wave,
    \vspace{-0.5 \baselineskip}
    \item[(vi)] and the cycle repeats.
\end{itemize}
\vspace{-0.5 \baselineskip}
Since the indirect term can efficiently emit density waves at the Lindblad resonances, it is easy to imagine that the wave would get stronger every cycle. But that would only induce an algebraic instability. To get a full exponential instability, one needs to close the feedback loop. The wave induces density perturbations, but we can usually pair an overdensity with an underdensity so that the gravitational pulls they exert on the central object cancel each other out. The only oscillation we cannot pair and cancel is the leftover one at the outer edge, so \cite{Shu+90} argue that this is where the central object's displacement is amplified every cycle.

This explanation suggests that SLING modes may be a sort of cavity modes that use another strategy to reflect the waves and create a cavity.

However, \cite{Noh+92} add that each time the wave refracts on the $Q$-barrier, part of it can tunnel through. This induces density waves in the inner disc (\textit{i.e.} below corotation), which also acts as a cavity. They show that in the inner cavity, we can get a process of SWING amplification \citep{Toomre81}, and that in `low-mass' discs (${ \Md / \Ms < 0.7}$), this process makes a significant contribution to the growth rate of the full instability.

\begin{figure*}
    \centering
    \vspace{-0.5 \baselineskip}
    \includegraphics[width = \linewidth]{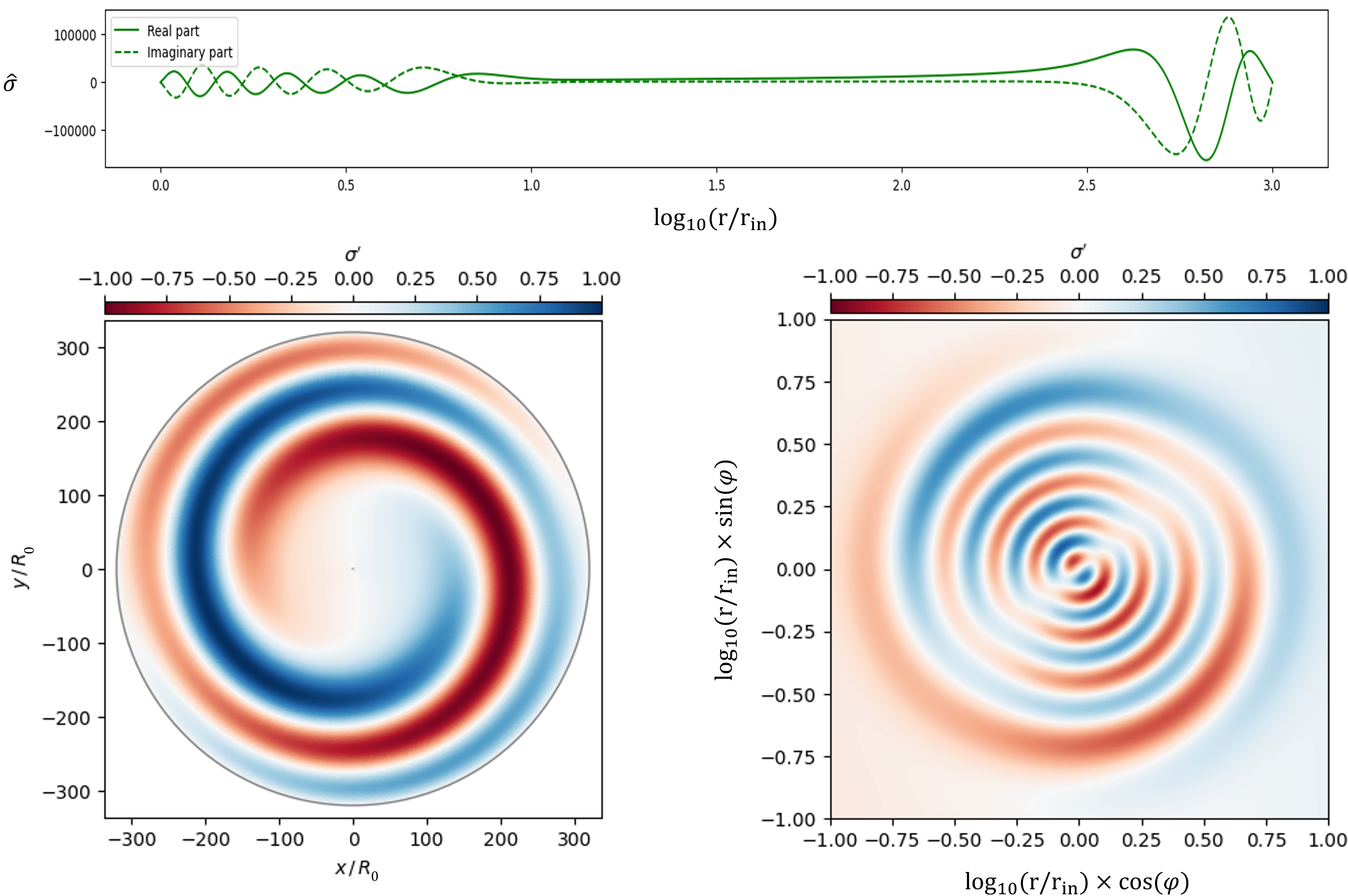}
    \caption{
        \textit{Top:} 1D profile of the relative density perturbation induced by a SLING mode. To produce this figure we used the disc model described in \S\ref{sec:Equilibrium}, but with \smash{${ \ro / R_{0} = 320 }$} and \smash{${ h = 0.01 }$}. We did find SLING modes with \smash{${ \ro / R_{0} = 3.125 }$}, but they induced small-scale oscillations that are hard to see on a PDF document. We did not, however, find any SLING mode with \smash{${h = 0.05 }$}.
        \textit{Bottom left:} Same data, but represented in 2D. We use a linear spatial scale to focus on the outer disc. The prominent ${ m = 1}$ spiral is indicative of density waves propagating outwards. We can see this in the top panel as well, from the ${ \pi / 2 }$ phase-shift between the real and imaginary parts of $\hs$. We also can see a hint of those density waves in Fig.~4 of \cite{Adams+89}.
        \textit{Bottom right:} Same data, but using a logarithmic spatial scale to focus on the inner disc. The deformation of the spiral is indicative of a balanced mix between inbound and outbound density waves. Once again, we can also see this in Fig.~4 of \cite{Adams+89} and in the top panel of the present figure, from the nearly perfect phase opposition between the real and imaginary parts of $\hs$. The unperturbed area between the inner disc and the outer disc is the $Q$-barrier.
    }
    \label{fig:SLING_modes}
    \vspace{-0.5 \baselineskip}
\end{figure*}

The quotation marks highlight that those publications were written at a time where direct observation of accretion discs had only just become feasible \citep{Mundy+86, Beckwith+86, SargentBeckwith87, KeeneMasson90}. This is why the parameters used by modellers were sometimes off by several orders of magnitude. For instance, the disc modelled by \cite{Adams+89} was as heavy as the central object. We show in Fig.~\ref{fig:SLING_modes} that we can reproduce the SLING instability in a `modern' part of parameter space (power-law discs, \smash{${ \Md / \Ms \leq 0.1}$}, \smash{${ \ro / \ri \leq 1000}$}), but only at low temperatures (\smash{${ h \leq 0.01}$}).

\subsection{Evanescent modes}
\label{sub:Gravitational_modes_evanescent}

\begin{figure*}
    \centering
    \includegraphics[width = \linewidth]{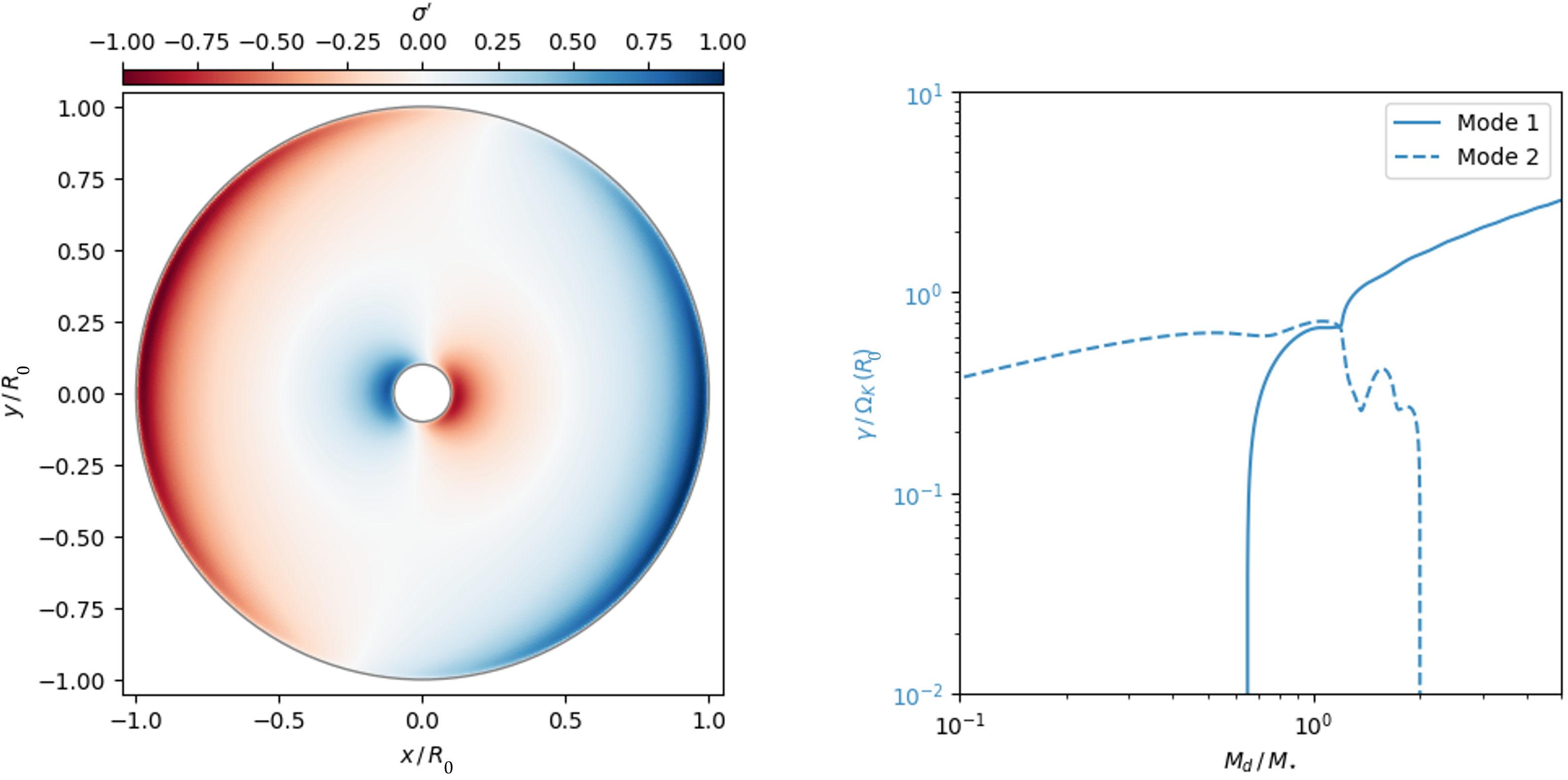}
    \caption{
        \textit{Left:} 2D representation of the relative density perturbations induced by an evanescent mode. Instead of the disc model described in \S\ref{sec:Equilibrium}, we use the polytropic (\smash{${ \gamma = 2 }$}) setup of \cite{Heemskerk+92}, except that we take \smash{${ h = 0.05 }$} and \smash{${ \Md / \Ms = 0.06 }$}. Notice how the mode is localised near the disc's boundaries.
        \textit{Right:} Growth rate of the evanescent mode as a function of the disc's mass. Apart from \smash{${ \Md / \Ms }$}, we use the same parameters as in the left panel. We use a continuation algorithm to follow two evanescent modes. They almost collide when \smash{${ \Md / \Ms \approx 1.2 }$}, so it is possible that the continuation algorithm mixed the two branches during the collision. This does not matter, the important point is that whatever the mass of the disc, there is always at least one fast-growing evanescent mode.
    }
    \label{fig:Evanescent_modes}
\end{figure*}

We propose this name for the modes discovered by \cite{Heemskerk+92} because they fit entirely within the $Q$-barrier, where density waves are evanescent. The first consequence of this is that evanescent modes can only induce perturbations close to the disc's boundaries, as shown in Fig.~\ref{fig:Evanescent_modes}-\textit{left}. The second consequence is that they can only arise in discs so narrow that the whole disc fits inside a $Q$-barrier. Astrophysical examples might include young decretion discs or AGN tori, but we find this unlikely.

\cite{Heemskerk+92} make a hand-wavy case against the existence of the instability in discs less massive than their central object, but we show in Fig.~\ref{fig:Evanescent_modes}-\textit{right} that this is incorrect. The evanescent instability is, however, fragile to modifications of the background surface density profile~\smash{$\oSigma$}. We did not manage to reproduce it in any power-law disc.

\section{Overdamped modes}
\label{sec:Overdamped_modes}

To exhibit the last group of modes, we must activate the damping layers. For the sake of simplicity, we shall also deactivate self-gravity. But the overdamped modes accommodate self-gravity without any difficulty.

\begin{figure*}
    \centering
    \includegraphics[width = \linewidth]{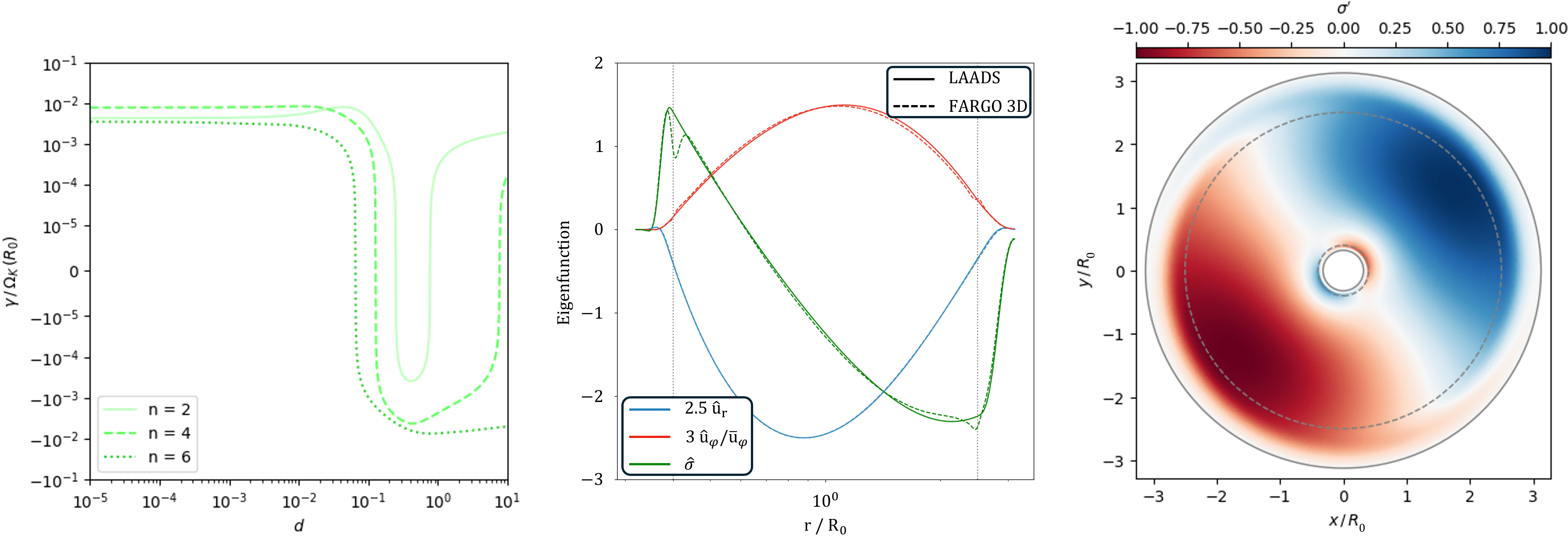}
    \caption{
        \textit{Left:} Growth rate $\gamma$ of the first three unstable cavity modes of Fig.~\ref{fig:Cavity_modes} as a function of the damping level~$d$. Notice that they are damped at intermediate damping levels, but that they become unstable again for high damping levels. This is why we call those modes `overdamped'.
        \textit{Centre:} Comparison between the eigenfunction of the \smash{${ n = 2 }$} overdamped mode predicted by LAADS and the structure of the fastest growing mode found in the Fargo3D experiments of \cite{Crida+25b}. The three variables are measured at three different azimuths~\smash{${ \varphi }$}, but we select the same three azimuths for LAADS and Fargo3D. The match is excellent everywhere except near the entrance of the damping layers (indicated by the dashed grey lines). The eigenvalues are also within 10\% of each other, so we conclude that the mode discovered by \cite{Crida+25b} is an overdamped mode.
        \textit{Right:} 2D representation of the relative density perturbations induced by the mode of the central panel. Once again, the dashed grey lines indicate the entrance of the damping layers.
    }
    \label{fig:Overdamped_modes}
\end{figure*}

As stated in \S\ref{sub:linear_perturbation_equations_damping_layers}, the point of the damping layers is to damp waves before they can reflect on the disc's boundaries. This tool is usually used to simulate planetary wakes, but here we hope that it will annihilate the cavity modes and the SLING modes, since both are caused by density wave reflection. As expected, Fig.~\ref{fig:Overdamped_modes}-\textit{left} shows that for intermediate damping levels, the cavity modes are damped. However, if one keeps increasing the damping level, they become unstable again -- hence why we call them `overdamped' modes. 

One may think that they can avoid reflex modes entirely by choosing their damping level carefully, but for some systems there is always either an unstable cavity mode or an overdamped mode, whatever the damping level. For instance, this is what happens in our fiducial disc when we work in the stellocentric reference frame and use Dirichlet \BCs\ on \smash{${ \hur }$}. Furthermore, if the damping layers are too weak, they cannot eliminate planetary~wakes.

The central panel of Fig.~\ref{fig:Overdamped_modes} shows that the unstable reflex mode discovered by \cite{Crida+25b} in their Fargo3D simulations is the ${ n = 2 }$ overdamped mode. This identification is one of the main results of the present paper. It also validates LAADS once more.

Finally, let us propose a tentative explanation for the origin of the overdamped modes. Since they only appear when the damping level is high, we suspect that they are cavity modes whose density waves reflect on the damping layers rather than the boundaries of the domain. Indeed, if the damping level is high, the properties of the medium in which the wave propagates change abruptly between the disc and the layer. And just like light waves can reflect on the boundary between air and water, density waves can reflect on the boundary between the disc and the damping layer.

That being said, Fig.~\ref{fig:Overdamped_modes}-\textit{right} and Fig.~\ref{fig:Cavity_modes}-\textit{top centre} show that while the ${ n = 2 }$ cavity mode induces spirals, the ${ n = 2 }$ overdamped mode creates crescents. This discrepancy may provide an argument against our explanation.

\section{Discussion and Conclusion}
\label{sec:Discussion_conclusion}

In this paper, we exhibited seven linear instabilities triggered by the reflex motion of the central object in response to non-axisymmetric density distributions in the disc. Four of those modes were already known: \cite{Sanchez+25} discovered the `bracket' modes, \cite{Adams+89} the SLING modes, \cite{Heemskerk+92} the `evanescent' modes, and \cite{Crida+25b} the `overdamped' modes. We contribute by checking whether those modes exist in modern sections of parameter space, showing that they are qualitatively different from one another, outlining their defining characteristics and proposing a classification. The other three `cavity' modes are, to the best of our knowledge, additions to the literature. 

All those modes are intimately linked to the boundaries of the simulation domain. Bracket modes only appear if the boundaries are inside the disc, evanescent modes only appear if the disc's boundaries are extremely close to each other and overdamped modes are caused by the damping layers, so those instabilities are artificial. The other modes depend on how density waves reflect at the boundaries, and wave reflection can be physical (at an ice line, at the edge of the gap carved by a planet, if the disc has a sharp boundary, ...), so they may be genuine in some astrophysical discs.

When those instabilities are real, they can have important consequences:
\begin{itemize}
    \vspace{-0.5 \baselineskip}
    \item[(i)] Unpublished non-linear SPH simulations by co-author JFG show that in some cases the disc explodes before reorganising into a new, lighter and eccentric disc. Reflex instabilities may therefore affect disc evolution, structure and dispersal.
    \vspace{-0.5 \baselineskip}
    \item[(ii)] \cite{Sanchez+25} pointed out that the bracket instability can drive disc eccentricity. This may well be true of the other instabilities, because several of them also induce \smash{${m = 1}$} spirals.
    \vspace{-0.5 \baselineskip}
    \item[(iii)] \cite{Sanchez+25} also pointed out that since the bracket instability induces density perturbations in the disc, it might affect the torque experienced by planets and therefore their migration. This is certainly true of all seven reflex instabilities.
    \vspace{-0.5 \baselineskip}
    \item[(iv)] We showed that reflex instabilities can induce one-armed spirals and large-scale crescents. As such, they might explain some of the substructures detected by ALMA in protoplanetary discs \citep{Bae+23} without invoking any unseen planet.
    \vspace{-0.5 \baselineskip}
    \item[(v)] Since the central object describes an outward spiral, prograde reflex instabilities must be funelling angular momentum from the disc to the central object. As such, they must drive accretion. Conversely, retrograde instabilities mitigate accretion.
\end{itemize}
\vspace{-0.5 \baselineskip}

But even when reflection is not physically justified, reflective boundary conditions will induce artificial reflex instabilities in many simulations. This will drive artificial disc eccentricity, bias planetary migration studies, create spurious substructures, cause artificial accretion, and may even make the disc explode. Since those artefacts are analytical rather than numerical, they will be hard to recognise. Since they depend on the reference frame, the surface density profiles, the boundary conditions, \textit{etc.} they will also be hard to predict. And since damping layers have an unfortunate tendency to create new instabilities, those artefacts will be hard to avoid. This leads us to suggest that our community may not have the right tools to model the reflex motion of the central object yet. We hope that our work will motivate the development of such a tool. In the meantime, it is safer to deactivate the indirect term \citep{Crida+25a}.

Our work also opens a wide range of questions about reflex instabilities. Indeed, we did not explain the physical mechanism of some instabilities, we do not know how they respond to additional physics like magnetohydrodynamics or thermodynamics, we need to study whether and how they generalise to 3D, and we are yet to study their non-linear outcome. Furthermore, our exploration of parameter space is far from exhaustive, so there are almost certainly other reflex instabilities beyond the seven described in the present paper. In fact, we find it surprising that the four degrees of freedom brought by the central object activate so many dynamical pathways.

Finally, the mathematical structure of our equations suggests that similar phenomena, questions and issues will arise in all systems where a disc is subject to a large-scale-and-slowly-evolving perturbation. This includes fly-bys, outer binary companions, photo-evaporation, ...

\section*{Acknowledgments}

NM acknowledges financial support from the Centre National d’\'Etudes Spatiales (CNES - France) within the framework of the Hera mission, from the COST action CA22133 `PLANETS', and from the French government through the \smash{${\text{UCA}^{\text{JEDI}}}$} `Investments in the Future' project managed by the National Research Agency (ANR) with the reference number \smash{ANR-15-IDEX-01}. All authors acknowledge financial support from the `Programme National Astronomie - Astrophysique' (PNAA) and `Programme National de Plan\'etologie' (PNP) of CNRS/INSU co-funded by CEA and CNES.

NM also wishes to thank G. Ogilvie, K. Kratter, A. Youdin and J. Li for their valuable insights.

\section*{Data Availability}

The data and numerical codes underlying this article were produced by the authors. They will be shared on reasonable request to the corresponding author.

\newpage
\section*{References}
\bibliographystyle{mnras}
\bibliography{main}

\appendix

\section{A. Analytical description of the cavity modes}
\label{sec:Analytical_description_of_the_cavity_modes}

In \S\ref{sec:Cavity_modes} we showed that the reflex motion of the central star activates three classes of cavity modes. The goal of the present appendix is to describe those analytically. The analytical machinery is heavy, so to keep things simple we shall only consider Dirichlet boundary conditions on $\hs$ in the stellocentric reference frame.

\subsection{A.1. General considerations}
\label{sub:Analytical_description_of_the_cavity_modes_general}

\subsubsection{A.1.1. Reduction of the system to a regular ODE}
\label{ssub:Analytical_description_of_the_cavity_modes_general_ODE}

\newcommand{\hphit}{\hat{\phi}_{t}} 

\renewcommand{\L}{\mathcal{L}} 

In the absence of self-gravity, we can introduce the total potential \smash{${ \hphit = \cs^{2} \hs + \hphii }$} and reduce the linear momentum equations~(\ref{eq:Fourier-transformed_equations_momentum_r}, \ref{eq:Fourier-transformed_equations_momentum_p}) to
\begin{align}
    i \C \, \hU - 2 \Omega \, \hV &= - \rd_{s} \hphit , \nonumber \\
    i \C \, \hV + \frac{\kappa^{2}}{2 \Omega} \, \hU &= i m \, \hphit , \nonumber
\end{align}
quickly leading, if ${ m = + 1 }$, to
\begin{subequations}
    \nonumber
    \begin{align}
        & \hU = i \, \frac{2 \Omega \, \hphit - \C \, \rd_{s} \hphit}{\L^{2}} , \nonumber \\
        & \hV = \frac{\frac{\kappa^{2}}{2 \Omega} \rd_{s} \hphit - \C \, \hphit}{\L^{2}} , \nonumber
    \end{align}
\end{subequations}
where ${ \L^{2} = \kappa^{2} - \C^{2} }$. We use the letter $\L$ to indicate that it is zero at the Lindblad resonances.

We can inject the previous formulas into the mass conservation equation~\eqref{eq:Fourier-transformed_equations_mass} to get a second-order \ODE\ on~$\hs$,
\begin{flalign}
    \!\!\!\!\!\!\!\!\!\!\!\!\!\!\!\!\!\!\!\!\!\!\!\!\!\!\!\!\! \cs^{2} \left[ \rd_{ss}^{2} \hs + a(r) \, \rd_{s} \hs + b(r) \hs \right] - r^{2} \L^{2} \, \hs \label{eq:Cavity_modes_ODE} \\
    \pushright{ \displaystyle = - \left[ \rd_{ss}^{2} \hphii + a(r) \, \rd_{s} \hphii + b(r) \hphii \right] .} 
\end{flalign}
where
\begin{equation}
    \nonumber
    \begin{cases}
        a(r) &= + \left[ \rd_{s} \ln{\left( \frac{\C \, \oSigma}{\L^{2}} \right) + \frac{1}{2} \frac{\kappa^{2}}{\Omega \C} - 2 \frac{\Omega}{\C}} \right] , \phantom{\displaystyle \frac{1}{1}} \\
        b(r) &= - \left[ 1 + 2 \frac{\Omega}{\C} \, \rd_{s} \ln{ \left( \frac{\Omega \oSigma}{\L^{2}} \right)} \right] , \phantom{\displaystyle \frac{1}{1}}
\end{cases}
\end{equation} \\

\newcommand{\F}{\mathcal{F}} 

Let us now adimensionalise the equations. To do so, we use a single reference lengthscale, \smash{$R_{0}$}, but two reference timescales: the complex frequency~\smash{$|\varpi|$} and the Keplerian frequency~\smash{${ \OmegaK (R_{0}) }$}. This is because we expect the \smash{${ r^{2} \L^{2} }$} term to act as a natural frequency, we would like its dimensionless form to be of order one, and we know that at low temperatures \smash{${ \L^{2} \sim 2 \, \varpi \, \Omega }$}. We therefore divide both sides of Eq.~\eqref{eq:Cavity_modes_ODE} by \smash{${ R_{0}^{2} \times \, |\varpi| \, \OmegaK (R_{0}) }$} to get
\begin{equation}
    \label{eq:Cavity_modes_ODE_v2}
    \varepsilon^{2} \, \rd_{ss}^{2} \hs + \varepsilon^{2} \, \nu (r) \, \rd_{s} \hs + k^{2}(r) \, \hs = \F(s) \, \hIT ,  
\end{equation}
where \smash{${ \varepsilon^{2} = h^{2} \times \OmegaK (R_{0}) / |\varpi| }$} and
\begin{equation}
    \nonumber
    \begin{cases}
        \nu (r) &= a(r) , \phantom{\displaystyle \frac{1}{1}} \\
        k^{2} (r) &= \left(\frac{r}{R_{0}}\right)^{2} \frac{- \L^{2} (r)}{\OmegaK (R_{0}) \, |\varpi|} + \varepsilon^{2} \, b(r) , \phantom{\displaystyle \frac{1}{1}} \\
        \F(s) &= \frac{r}{R_{0}} \frac{\OmegaK (R_{0})}{|\varpi|} \, \left[ 1 + a(r) + b(r) \right] . \phantom{\displaystyle \frac{1}{1}}
    \end{cases}
\end{equation}

We shall focus on low-frequency modes in cold discs, such that \smash{${ h^{2} \ll |\varpi| / \OmegaK (R_{0}) \ll 1 }$}. This ensures that $\varepsilon$ is much smaller than one and that \smash{${ k^{2} }$} is of order one. We shall also focus on retrograde modes to make sure that \smash{${ k^{2} }$} has a positive real part.

\subsubsection{A.1.2. WKB analysis}
\label{ssub:Analytical_description_of_the_cavity_modes_general_WKB}

\newcommand{\hS}{\hat{S}} 

\newcommand{\si}{s_{\text{in}}} 
\renewcommand{\so}{s_{\text{out}}} 

To approximate the solutions to Eq.~\eqref{eq:Cavity_modes_ODE_v2}, we use WKB analysis \citep{BenderOrszag99}. We start by writing the perturbed density as an exponential power series,
\begin{equation}
    \label{eq:WKB_ansatz}
    \hs (s) = \exp{\left[ \frac{1}{\varepsilon} \sum_{\ell = 0}^{\infty} \varepsilon^{\ell} \hS_{\ell}(s) \right]} .
\end{equation}
Injecting this into the homogeneous equation associated to Eq.~\eqref{eq:Cavity_modes_ODE_v2} and grouping the terms by powers of $\varepsilon$ gives a sequence of equations,
\begin{align}
    & (\rd_{s} \hS_{0})^{2} + k^{2} = 0 , \nonumber \\
    & \forall \ell \! \geq \! 1, \,\, \rd_{ss}^{2} \hS_{\ell - 1} + \nu \, \rd_{s} \hS_{\ell - 1} + \! \sum_{m = 0}^{\ell} (\rd_{s} \hS_{m}) (\rd_{s} \hS_{\ell - m}) = 0 , \nonumber
\end{align}
which can be solved successively. The first two give
\begin{align}
    \hS_{0} (s) & = \pm i \int_{\si}^{s} \!\!\! \rd s' \, k(s') , \nonumber \\
    \hS_{1} (s) & = - \frac{1}{2} \left\{ \ln{\left[ k (s) \right]} + \int_{\si}^{s} \!\!\! \rd s' \, \nu(s') \right\} ,\nonumber
\end{align}
with \smash{${ \si = \ln{(\ri / R_{0})} }$}. This leads to
\begin{small}
\begin{equation}
    \label{eq:WKB_solution_homogneous}
    \!\!\! \hs_{\pm} (s) = k^{-1/2} (s) \times \exp{\left\{ \pm \, \frac{i}{\varepsilon} \!\! \int_{\si}^{s} \!\!\!\!\! \rd s' \, k (s') \! - \! \frac{1}{2} \!\! \int_{\si}^{s} \!\!\!\!\! \rd s' \, \nu (s') \! \right\}} . \!\!\!\!
\end{equation}
\end{small}
\!\!\!\!Note that when \smash{${ k^{2} }$} is a complex number, we define \smash{${ k }$} as the root of \smash{${ k^{2} }$} that has a positive real part, and \smash{${ k^{1/2} }$} as the root of \smash{${ k }$} that has a positive real part.

\begin{figure*}
    \centering
    \includegraphics[width = \linewidth]{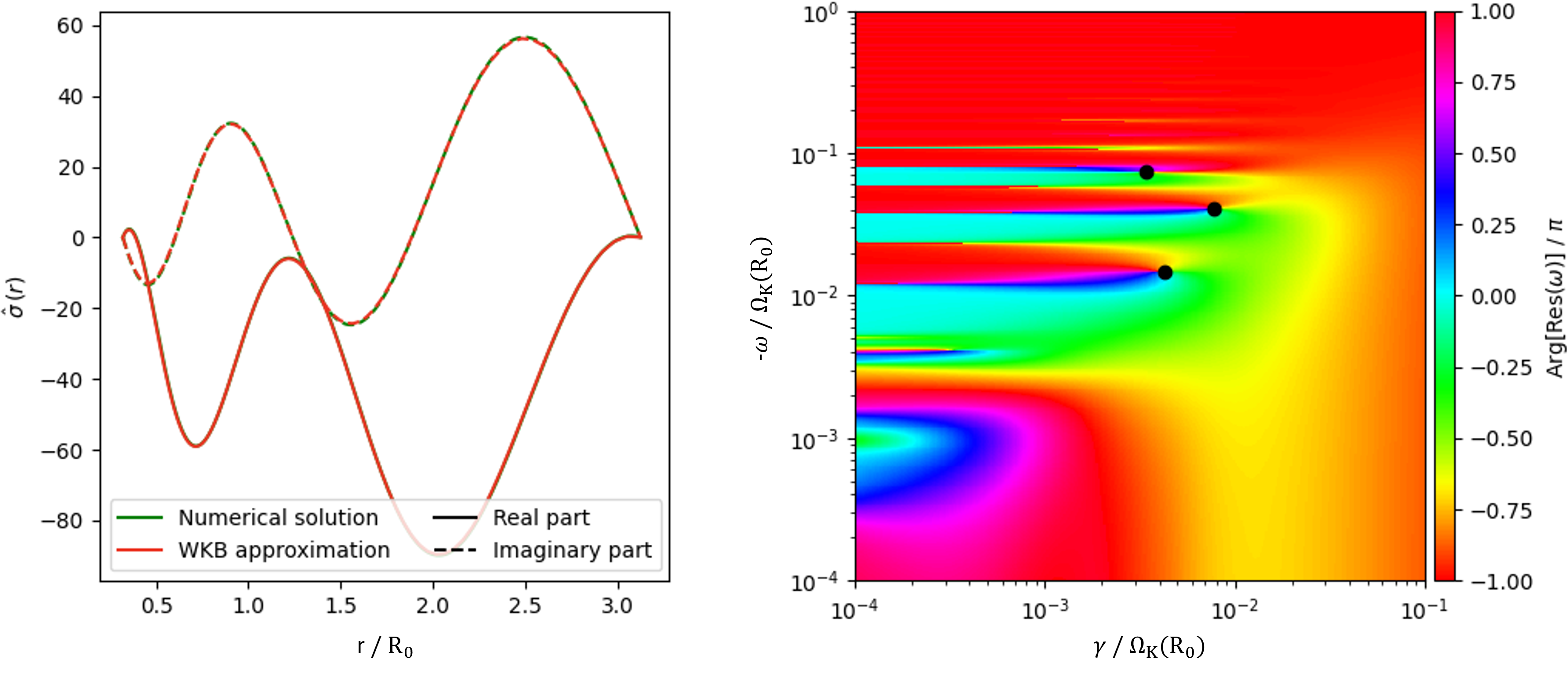}
    \caption{
        \textit{Left:} Comparison between the density perturbations predicted by our semi-analytical theory (\textit{i.e.} Eqs.~\ref{eq:WKB_solution_general} and~\ref{eq:Cavity_modes_standard_lambda}) and by our code LAADS. The agreement is so good that the red and green curves are undistinguishable.
        \textit{Right:} Map of the residual \smash{${ \hIT - \int_{\si}^{\so} \!\! \rd s \, w(s) \, \hs (s) }$} left in Eq.~\eqref{eq:WKB_loop_closing_equation} by different complex frequencies \smash{${ \varpi = \omega - i \gamma }$}. Cauchy's principle indicates that the argument of the residual should loop once or several times around each of the system's eigenfrequencies. We clearly see three such points (the fourth potential point in \smash{${ \gamma / \OmegaK (R_{0}) = 4 \times 10^{-4} \text{ and } \omega / \OmegaK (R_{0}) = -4 \times 10^{-3} }$} is probably a resolution artefact). We overlaid black dots to indicate the eigenfrequencies found by LAADS. There are indeed three, exactly where predicted by our semi-analytical theory. Those two plots are strong validations both of our analytical work and of our code. Indeed, neither approach offers any free parameter, so the match between the two approaches is not due to a fitting procedure.
        }
        \label{fig:WKB_analysis_vs_LAADS}
\end{figure*}

\newcommand{\hsp}{\hs_{\text{p.}}} 
\newcommand{\hcS}{\hat{\mathcal{S}}} 

Equation~\eqref{eq:WKB_solution_homogneous} approximates the homogeneous solutions to Eq.~\eqref{eq:Cavity_modes_ODE_v2}. To approximate a particular solution, we use the variation of parameters. We look for a solution of the form
\begin{equation}
    \nonumber
    \hsp (s) = \hcS_{+} (s) \, \hs_{+} (s) + \hcS_{-} (s) \, \hs_{-} (s) .
\end{equation}
Generally, we have 
\begin{equation}
    \nonumber
    \rd_{s} \hsp = \hs_{+} \, \rd_{s} \hcS_{+} + \hs_{-} \, \rd_{s} \hcS_{-} + \hcS_{+} \, \rd_{s} \hs_{+} + \hcS_{-} \, \rd_{s} \hs_{-} ,
\end{equation}
but since \smash{${\hs_{-}}$} is never null, we can replace \smash{${ (\hcS_{+}, \hcS_{-}) }$} by 
\begin{equation}
    \nonumber
    (\hcS_{+}, \hcS_{-}) \rightarrow (\hcS'_{+} = \hcS_{+} + \Delta \hcS, \,\,\, \hcS'_{-} = \hcS_{-} - [\hs_{+} / \hs_{-}] \, \Delta \hcS)
\end{equation}
to get
\begin{small}
\begin{equation}
    \nonumber
    \hs_{+} \, \rd_{s} \hcS'_{+} + \hs_{-} \, \rd_{s} \hcS'_{-} = \hs_{+} \, \rd_{s} \hcS_{+} + \hs_{-} \, \rd_{s} \hcS_{-} -  \hs_{-} \, \rd_{s} \left[ \hs_{+} / \hs_{-} \right] \Delta \hcS .
\end{equation}
\end{small}
\!\!\!\!\!Now \smash{${ \hs_{-} \, \rd_{s} \left[ \hs_{+} / \hs_{-} \right] }$} is never null, so we can choose ${ \Delta \hcS }$ such that \smash{${ \hs_{+} \rd_{s} \hcS'_{+} + \hs_{-} \rd_{s} \hcS'_{-} = 0 }$}. We can therefore impose the gauge condition
\begin{equation}
    \label{eq:WKB_gauge_condition}
    \hs_{+} \, \rd_{s} \hcS_{+} + \hs_{-} \, \rd_{s} \hcS_{-} = 0
\end{equation}
without loss of generality. We then have
\begin{equation}
    \nonumber
    \rd_{ss}^{2} \hsp = \rd_{s} \hcS_{+} \, \rd_{s} \hs_{+} + \rd_{s} \hcS_{-} \, \rd_{s} \hs_{-} + \hcS_{+} \, \rd_{ss}^{2} \hs_{+} + \hcS_{-} \, \rd_{ss}^{2} \hs_{-} , 
\end{equation}
and therefore
\begin{equation}
    \label{eq:WKB_equation_at_the_end_of_the_variation_of_parameters}
    \varepsilon^{2} \left[ \rd_{s} \hcS_{+} \, \rd_{s} \hs_{+} + \rd_{s} \hcS_{-} \, \rd_{s} \hs_{-} \right] = \F(s) \, \hIT.
\end{equation}
We can combine Eqs.~\eqref{eq:WKB_gauge_condition} and~\eqref{eq:WKB_equation_at_the_end_of_the_variation_of_parameters} to get
\begin{equation}
    \nonumber
    \rd_{s} \hcS_{\pm} = \mp \frac{i}{2 \varepsilon} \, \hs_{\mp} (s) \, \F(s) \, \exp{\left[ \int_{\si}^{s} \!\!\!\!\! \rd s' \, \nu (s') \right]} \, \hIT .
\end{equation} \\

\noindent This leads to find the particular solution
\begin{small}
\begin{flalign}
    \!\!\!\!\! \hsp (s) \! = \! \frac{-i}{2 \varepsilon} \bigg\{ \hs_{+} (s) \! \times \! \int_{\si}^{s} \!\!\!\! \rd s' \, \hs_{-} (s') \, \F(s') \exp{\big[ \int_{\si}^{s'} \!\!\!\!\! \rd s'' \, \nu (s'') \big]} \nonumber \\
    \pushright{- \, \hs_{-} (s) \! \times \! \displaystyle \int_{\si}^{s} \!\!\!\! \rd s' \, \hs_{+} (s') \, \F(s') \exp{\big[ \int_{\si}^{s'} \!\!\!\!\! \rd s'' \, \nu (s'') \big]} \bigg\}  \, \hIT ,} \nonumber
\end{flalign}
\end{small}
\!\!\!which we can approximate using repeated integration by parts by
\begin{equation}
    \label{eq:WKB_particular_solution}
    \hsp (s) \approx \left[ \frac{\F (s)}{k^{2} (s)} - \frac{\hs_{+} (s) + \hs_{-} (s)}{2} \, \frac{\F (\si)}{k^{3/2} (\si)} \right] \hIT .
\end{equation}

\textit{In fine}, the general solution is
\begin{equation} 
    \label{eq:WKB_solution_general}
    \hs (s) = \hsp (s) + \lambda_{+} \, \hs_{+} (s) + \lambda_{-} \, \hs_{-} (s) .
\end{equation}
Naturally, the integration constants \smash{${ \lambda_{+} }$} and \smash{${ \lambda_{-} }$} depend on the boundary conditions. We explain how to estimate them in \S\hyperref[sub:Analytical_description_of_the_cavity_modes_standard]{A.2} and~\hyperref[sub:Analytical_description_of_the_cavity_modes_quasi-resonant]{A.4}. 

For now, let us simply point at Fig.~\ref{fig:WKB_analysis_vs_LAADS}-\textit{left} to show that Eq.~\eqref{eq:WKB_solution_general} provides an extremely precise description of the cavity eigenmodes, even when \smash{${ \varepsilon }$} is of order 0.1.

\subsubsection{A.1.3. Approximating the indirect potential}
\label{ssub:ssub:Analytical_description_of_the_cavity_modes_general_potential}

Once we know the density perturbation caused by a potential perturbation, we need to close the loop and compute the potential perturbation caused by that density perturbation. This comes down to the loop-closing equation
\begin{equation}
    \label{eq:WKB_loop_closing_equation}
    \hIT = \int_{\si}^{\so} \!\!\! \rd s \, w(s) \, \hs (s) ,
\end{equation}
where \smash{${ w (s)  \! = \! - \delta_{m}^{\pm 1} \pi G \oSigma (r) / [ R_{0} \, \OmegaK^{2} (R_{0}) ] }$} is a slowly-vary- ing `weight' function deduced from Eqs.~\eqref{eq:Fourier_transformed_equations_indirect_potential}~and~\eqref{eq:Fourier_transformed_equations_indirect_term}.

We can decompose the integral \smash{$I$} into three parts,
\begin{flalign}
    \nonumber
    \!\!\!\!\!\!\!\!\!\!\!\!\!\!\!\!\! I = \int_{\si}^{\so} \!\!\!\! \rd s \, w(s) \frac{\F(s) \, \hIT}{k^{2} (s)} + \left(\lambda_{+} - \frac{\F(\si) \, \hIT}{2 \, k^{3/2} (\si)}\right) I_{+} \nonumber \\
    \pushright{\displaystyle + \left(\lambda_{-} - \frac{\F(\si) \, \hIT}{2 \, k^{3/2} (\si)}\right) I_{-} ,} \nonumber
\end{flalign}
where \smash{${ I_{\pm} = \int_{\si}^{\so} \rd s \, w(s) \, \hs_{\pm} (s) }$}.

Those two integrals are hard to approximate. Indeed, \smash{${ k }$} is complex, so we expect the integrand to both oscillate and explode. We could deal with oscillations alone using repeated integration by part, or with explosions alone using Laplace's method. Unfortunately, there is no textbook method to deal with both at the same time, so we had to devise a tailored strategy. We start by introducing
\begin{equation}
    \nonumber
    I_{\pm} (s) = \int_{\si}^{s} \!\!\! \rd s' \, f(s) \, \exp{\left[ \pm \, \frac{i}{\varepsilon} \!\! \int_{\si}^{s'} \!\!\!\!\! \rd s'' \, k (s'') \right]}
\end{equation}
with \smash{${ f(s) \! = \! w(s) \, k^{-1/2} (s) \, \exp{[ - \frac{1}{2} \int_{\si}^{s} \! \rd s' \, \nu(s') ]} }$}. We then introduce the function ${ J_{\pm} }$ such that
\begin{equation}
    \nonumber
    I_{\pm} (s) = J_{\pm} (s) \, \exp{\left[ \pm \, \frac{i}{\varepsilon} \!\! \int_{\si}^{s} \!\!\!\!\! \rd s' \, k (s') \right]} .
\end{equation}
This new function satisfies
\begin{equation}
    \label{eq:Approximate_the_integral_ODE_on_J}
    \varepsilon \, \rd_{s} J_{\pm} \pm i k(s) \, J_{\pm} (s) = \varepsilon \, f(s) .
\end{equation}

We tried to approximate a particular solution to Eq.~\eqref{eq:Approximate_the_integral_ODE_on_J} using a power series, but it converged slowly. So instead, we approximate it using continued fractions. Specifically, we construct a sequence of approximations
\begin{equation}
    \nonumber
    J_{1}^{\pm} = \varepsilon \, j_{1}^{\pm}, \quad J_{2}^{\pm} = \frac{\varepsilon \, j_{1}^{\pm}}{1 + \varepsilon \, j_{2}^{\pm}}, \quad J_{3}^{\pm} = \frac{\varepsilon \, j_{1}^{\pm}}{1 + \frac{\varepsilon \, j_{2}^{\pm}}{1 + \varepsilon \, j_{3}^{\pm}}}, \quad \text{...}
\end{equation}
where the \smash{${ j_{\ell}^{\pm} }$} are carefully-chosen functions. For instance, the first two are
\begin{equation}
    \nonumber
    \begin{cases}
        j_{1}^{\pm} (s) &= \mp i \, f(s) / k(s) , \phantom{\displaystyle \frac{A}{A}} \\
        j_{2}^{\pm} (s) &= \mp i \, \rd_{s} \! \ln{[f(s) / k(s)]} \, / \, k(s) . \phantom{\displaystyle \frac{A}{A}}
    \end{cases}
\end{equation}
If we stop the expansion at this order, we get
\begin{equation}
    \label{eq:Approximate_the_integral_order_2}
    I_{\pm} = \mp \bigg[ \frac{i \varepsilon \, w(s) \, \hs_{\pm}(s)}{k(s) \mp i \varepsilon \, \rd_{s} \ln{[ f(s) / k(s) ]}} \bigg]_{\si}^{\so} \!\!\!\!\!\!\!\!\! . 
\end{equation}
In all the cases we tested, this was accurate to within 5\% even when $\varepsilon$ was of order 0.1. By comparison, the first-order expansion was only accurate to within 15\%, and the second-order power series expansion to within 10\%.

\subsection{A.2. Standard cavity modes}
\label{sub:Analytical_description_of_the_cavity_modes_standard}

At this stage, all that is left to do is to estimate the integration constants \smash{$\lambda_{+}$} and \smash{$\lambda_{-}$}. Since \smash{${ \hs_{+} (\si) = \hs_{-} (\si) }$}, we find that \smash{${ \hsp(\si) = 0}$}. Therefore, the inner Dirichlet boundary condition on $\hs$ gives \smash{${ \lambda_{-} = - \lambda_{+} }$}. This much is true for all cavity modes.

For the standard cavity modes, \smash{${ \int_{\si}^{\so} \! \rd s \, k(s) }$} is far from being a multiple of \smash{${ \varepsilon \pi }$} so \smash{${ \hs_{+} (\so) \neq \hs_{-}(\so) }$} and the outer boundary condition gives 
\begin{equation}
    \label{eq:Cavity_modes_standard_lambda}
    \lambda_{+} = - \frac{\hsp (\so)}{\hs_{+} ( \so) - \hs_{-} (\so)} .
\end{equation}

\subsection{A.3. Resonant cavity modes}
\label{sub:Analytical_description_of_the_cavity_modes_resonant}

\noindent If the resonance condition
\begin{equation}
    \label{eq:Cavity_modes_resonance_condition}
    \int_{\si}^{\so} \!\!\!\!\! \rd s \, \frac{k (s)}{\varepsilon} =  n \pi, \,\,\, n \in \mathbb{N}^{*}
\end{equation}
is satisfied, the outer boundary condition becomes
\begin{small}
\begin{equation}
    \nonumber
    \frac{\F (\so)}{k^{2} (\so)} \! = \! (-1)^{n} \, k^{-1/2} (\so) \, \exp{\left\{ \! - \frac{1}{2} \int_{\si}^{\so} \!\!\!\!\!\!\! \rd s \, \nu (s) \right\}} \, \frac{\F (\si)}{k^{3/2} (\si)} .
\end{equation}
\end{small}
\!\!\!\!This equation is not satisfied. To understand why, notice in Eq.~\eqref{eq:Cavity_modes_standard_lambda} that \smash{${ \lambda_{+} }$} goes to infinity as we approach resonance. This is a sign that the response of the system at the resonant frequency is not a Fourier mode.

\subsection{A.4. Quasi-resonant cavity modes}
\label{sub:Analytical_description_of_the_cavity_modes_quasi-resonant}

If a cavity mode falls close to resonance, then Eq.~\eqref{eq:Cavity_modes_standard_lambda} becomes impractical. Indeed, it is only semi-analytical because of the $k$ and $\nu$ integrals in \smash{${ \hs_{\pm} }$}, and one would need high numerical resolution to evaluate those integrals at the precision required for \smash{${ 1 / [\hs_{+} (\so) - \hs_{-} (\so)] }$} to be accurate.

It is much easier to estimate \smash{${ \lambda_{+} }$} through the loop-closing equation~\eqref{eq:WKB_loop_closing_equation}:
\begin{small}
\begin{equation}
    \label{eq:Cavity_modes_resonant_lambda}
    \lambda_{+} = \frac{\displaystyle 1 - \int_{\si}^{\so} \!\!\!\! \rd s \, \frac{w(s) \, \F(s)}{k^{2} (s)} + \frac{\F(\si)}{2 \, k^{3/2} (\si)} \left(I_{+} + I_{-}\right)}{I_{+} - I_{-}} .
\end{equation}
\end{small} \\


\newpage
\onecolumngrid
\section{B. Parameters used in the figures}
\label{sec:Parameters}

We showed a lot of figures in the main part of the paper without specifying the exact values of the parameters used to produce them. The point of the present appendix is to correct this deficiency and make our work easy to reproduce.

\subsection{Figure~\ref{fig:Bracket_modes}}

\noindent \textit{All panels:} ${ p = 1.5 }$, ${ \ri / R_{0} = 0.32 }$, ${ \ro / R_{0} = 3.125 }$, ${ \Md / \Ms = 0.06 }$. \vspace{+0.5 \baselineskip}

\noindent $\bullet$ \textit{Top left}: Stellocentric reference frame, Neumann \BCs\ on \smash{$\hup$}, ${ h }$ variable, \smash{${ \omega / \OmegaK (R_{0}) = - 5 \times 10^{-2} }$}, \smash{${ \gamma / \OmegaK (R_{0}) = 0 }$}.

\noindent $\bullet$ \textit{Top right}: Barycentric reference frame, Dirichlet \BCs\ on \smash{$\hs$}, ${ h }$ variable, \smash{${ \omega }$} and \smash{${ \gamma }$} from Fig.~\ref{fig:Bracket_modes}-\textit{bottom left}.

\noindent $\bullet$ \textit{Bottom left}: Barycentric reference frame, Dirichlet \BCs\ on \smash{$\hs$}, ${ h }$ variable, \smash{${ \omega }$} and \smash{${ \gamma }$} variable.

\noindent $\bullet$ \textit{Bottom right}: Barycentric reference frame, Dirichlet \BCs\ on \smash{$\hs$}, ${ h = 0.05 }$, \smash{${ \omega / \OmegaK (R_{0}) = - 3.7 \times 10^{-2} }$}, $ \gamma / \OmegaK (R_{0}) = 4.4 \times 10^{-1} $, normalised such that \smash{${ \hat{r}_{\star, x} / R_{0} = 1.0 }$}.

\subsection{Figure~\ref{fig:Collisional_modes}}

\noindent \textit{All panels:} Stellocentric reference frame, Dirichlet \BCs\ on \smash{$\hs$}, ${ p = 1.5 }$, ${ \ri / R_{0} = 0.32 }$, ${ \ro / R_{0} = 3.125 }$, ${ \Md / \Ms = 0.06 }$, normalised such that \smash{${ \hIT = 1.0 }$}. \vspace{+0.5 \baselineskip}

\noindent $\bullet$ \textit{Top centre}: ${ h = 0.02 }$, \smash{${ \omega / \OmegaK (R_{0}) = - 1.5 \times 10^{-2} }$}, \smash{${ \gamma / \OmegaK (R_{0}) = 0 }$}.

\noindent $\bullet$ \textit{Top right}: ${ h = 0.0224 }$, \smash{${ \omega / \OmegaK (R_{0}) = - 2.0 \times 10^{-2} }$}, \smash{${ \gamma / \OmegaK (R_{0}) = 1.0 \times 10^{-4} }$}.

\noindent $\bullet$ \textit{Bottom left}: ${ h = 0.04 }$, \smash{${ \omega / \OmegaK (R_{0}) = - 5.1 \times 10^{-2} }$}, \smash{${ \gamma / \OmegaK (R_{0}) = 6.0 \times 10^{-3} }$}.

\noindent $\bullet$ \textit{Bottom centre}: ${ h = 0.0514 }$, \smash{${ \omega / \OmegaK (R_{0}) = - 7.7 \times 10^{-2} }$}, \smash{${ \gamma / \OmegaK (R_{0}) = 7.9 \times 10^{-4} }$}.

\noindent $\bullet$ \textit{Bottom right}: ${ h = 0.0515 }$, \smash{${ \omega / \OmegaK (R_{0}) = - 7.7 \times 10^{-2} }$}, \smash{${ \gamma / \OmegaK (R_{0}) = 0 }$}.

\subsection{Figure~\ref{fig:Cavity_modes}}

\noindent \textit{All panels:} Stellocentric reference frame, Dirichlet \BCs\ on \smash{$\hs$}, ${ p = 1.5 }$, ${ \ri / R_{0} = 0.32 }$, ${ \ro / R_{0} = 3.125 }$, ${ \Md / \Ms = 0.06 }$. \vspace{+0.5 \baselineskip}

\noindent $\bullet$ \textit{Left}: ${ h }$ variable, \smash{${ \omega }$} and \smash{${ \gamma }$} variable.

\noindent $\bullet$ \textit{Top centre}: ${ h = 0.05 }$, \smash{${ \omega / \OmegaK (R_{0}) = - 1.5 \times 10^{-2} }$}, \smash{${ \gamma / \OmegaK (R_{0}) = 4.3 \times 10^{-3} }$}, normalised such that \smash{${ \hIT = 1.0 }$}.

\noindent $\bullet$ \textit{Top right}: ${ h = 0.05 }$, \smash{${ \omega / \OmegaK (R_{0}) = - 4.1 \times 10^{-2} }$}, \smash{${ \gamma / \OmegaK (R_{0}) = 7.7 \times 10^{-3} }$}, normalised such that \smash{${ \hIT = 1.0 }$}.

\noindent $\bullet$ \textit{Bottom centre}: ${ h = 0.05 }$, \smash{${ \omega / \OmegaK (R_{0}) = - 7.4 \times 10^{-2} }$}, \smash{${ \gamma / \OmegaK (R_{0}) = 3.4 \times 10^{-3} }$}, normalised such that \smash{${ \hIT = 1.0 }$}.

\noindent $\bullet$ \textit{Bottom right}: ${ h = 0.03 }$, \smash{${ \omega / \OmegaK (R_{0}) = - 5.2 \times 10^{-2} }$}, \smash{${ \gamma / \OmegaK (R_{0}) = 5.1 \times 10^{-3} }$}, normalised such that \smash{${ \hIT = 1.0 }$}.

\subsection{Figure~\ref{fig:SLING_modes}}

\noindent \textit{All panels:} Stellocentric reference frame, Dirichlet \BCs\ on \smash{$\hs$}, ${ p = 1.5 }$, ${ \ri / R_{0} = 0.32 }$, ${ \ro / R_{0} = 320 }$, ${ \Md / \Ms = 0.06 }$, ${ h = 0.01 }$, \smash{${ \omega / \OmegaK (R_{0}) = 8.3 \times 10^{-4} }$}, \smash{${ \gamma / \OmegaK (R_{0}) = 1.4 \times 10^{-4} }$}, self-gravity on, \smash{${ \epsilon = 10^{-2} }$}, normalised such that \smash{${ \hIT = 1.0 }$}.

\subsection{Figure~\ref{fig:Evanescent_modes}}

\noindent \textit{All panels:} Surface density profile from \cite{Heemskerk+92}, temperature profile in \smash{${ r^{-1} }$}, polytropic equation of state with \smash{${ \gamma = 2 }$}, barycentric reference frame, Dirichlet \BCs\ on \smash{$\hs$}, ${ \ri / R_{0} = 0.1 }$, ${ \ro / R_{0} = 1.0 }$, ${ h = 0.05 }$, self-gravity on, \smash{${ \epsilon = 10^{-2} }$}. \vspace{+0.5 \baselineskip}

\noindent $\bullet$ \textit{Left}: ${ \Md / \Ms = 0.06 }$, \smash{${ \omega / \OmegaK (R_{0}) = 2.0 \times 10^{-2} }$}, \smash{${ \gamma / \OmegaK (R_{0}) = 3.0 \times 10^{-1} }$}, normalised such that \smash{${ \hat{r}_{\star, x} / R_{0} = 1.0 }$}.

\noindent $\bullet$ \textit{Right}: ${ \Md / \Ms }$ variable, \smash{${ \omega }$} and \smash{${ \gamma }$} variable.

\subsection{Figure~\ref{fig:Overdamped_modes}}

\noindent \textit{All panels:} Stellocentric reference frame, Dirichlet \BCs\ on \smash{$\hs$}, ${ p = 1.5 }$, ${ \ri / R_{0} = 0.32 }$, ${ \ro / R_{0} = 3.125 }$, ${ \Md / \Ms = 0.06 }$, ${ h = 0.05 }$, damping layers on. \vspace{+0.5 \baselineskip}

\noindent $\bullet$ \textit{Left}: ${ d }$ variable, \smash{${ \omega }$} and \smash{${ \gamma }$} variable.

\noindent $\bullet$ \textit{Centre}: ${ d  = 3.33 }$, \smash{${ \omega / \OmegaK (R_{0}) = - 2.9 \times 10^{-3} }$}, \smash{${ \gamma / \OmegaK (R_{0}) = 1.2 \times 10^{-3} }$}, normalised such that \smash{${ \hur = 1 + 0 i}$} at the radius where \smash{${ |\hur (r) | }$} is maximal, \smash{$\hur$} measured at an azimuth of \smash{${ \varphi = \pi }$}, \smash{${ \hup }$} and \smash{${ \hs }$} measured at an azimuth of \smash{${ \varphi = - \pi / 2 }$}, renormalised such that \smash{${ \max (|u_{r}'|) = 1.0 }$}.

\noindent $\bullet$ \textit{Right}: ${ d  = 3.33 }$, \smash{${ \omega / \OmegaK (R_{0}) = - 2.9 \times 10^{-3} }$}, \smash{${ \gamma / \OmegaK (R_{0}) = 1.2 \times 10^{-3} }$}, normalised such that \smash{${ \hIT = 1.0 }$}.

\subsection{Figure~\ref{fig:WKB_analysis_vs_LAADS}}

\noindent \textit{All panels:} Stellocentric reference frame, Dirichlet \BCs\ on \smash{$\hs$}, ${ p = 1.5 }$, ${ \ri / R_{0} = 0.32 }$, ${ \ro / R_{0} = 3.125 }$, ${ \Md / \Ms = 0.06 }$, ${ h = 0.05 }$, normalised such that \smash{${ \hIT = 1.0 }$} \vspace{+0.5 \baselineskip}

\noindent $\bullet$ \textit{Left}: \smash{${ \omega / \OmegaK (R_{0}) = - 4.1 \times 10^{-2} }$}, \smash{${ \gamma / \OmegaK (R_{0}) = 7.7 \times 10^{-3} }$}.

\noindent $\bullet$ \textit{Right}: \smash{${ \omega }$} and \smash{${ \gamma }$} variable.

\vspace{+1 \baselineskip}

\end{document}